\documentclass[reprint,superscriptaddress,amsmath,amssymb,aps,prl]{revtex4-2}

\usepackage{graphicx}
\usepackage{dcolumn}
\usepackage{bm}
\usepackage{color}
\usepackage[normalem]{ulem}
\usepackage{hyperref}

\begin{document}

\preprint{APS/123-QED}

\title{Continuum of magnetic excitations in the Kitaev honeycomb iridate D$_3$LiIr$_2$O$_6$}
\author{Thomas Halloran}
\affiliation{Institute for Quantum Matter and Department of Physics and Astronomy, Johns Hopkins University, Baltimore MD 21218, USA}
\affiliation{NIST Center for Neutron Research, Gaithersburg, Maryland\ 20899, USA}
\affiliation{Department of Physics and Astronomy, University of Maryland}
\author{Yishu Wang}
\affiliation{Institute for Quantum Matter and Department of Physics and Astronomy, Johns Hopkins University, Baltimore MD 21218, USA}
\affiliation{Department of Materials Science and Engineering, The University of Tennessee Knoxville}
\affiliation{Department of Physics and Astronomy, The University of Tennessee Knoxville}
\author{K.W. Plumb}
\affiliation{ Department of Physics, Brown\ University}
\author{M.B. Stone}
\affiliation{Neutron Scattering Division, Oak\ Ridge\ National\ Laboratory,\ Oak\ Ridge,\ Tennessee\ 37831, USA}
\author{Barry Winn}
\affiliation{Neutron Scattering Division, Oak\ Ridge\ National\ Laboratory,\ Oak\ Ridge,\ Tennessee\ 37831, USA}
\author{M. K. Graves-Brook}
\affiliation{Neutron Scattering Division, Oak\ Ridge\ National\ Laboratory,\ Oak\ Ridge,\ Tennessee\ 37831, USA}
\author{J. A. Rodriguez-Rivera}
\affiliation{NIST Center for Neutron Research, Gaithersburg, Maryland\ 20899, USA}
\affiliation{Department of Materials Science, University of Maryland, College Park, MD 20742}
\author{Yiming Qiu}
\affiliation{NIST Center for Neutron Research, Gaithersburg, Maryland\ 20899, USA}
\author{Prashant Chauhan}
\affiliation{Institute for Quantum Matter and Department of Physics and Astronomy, Johns Hopkins University, Baltimore MD 21218, USA}
\author{Johannes Knolle}
\affiliation{Department of Physics TQM, Technische Universität München, James-Franck-Straße 1, D-85748 Garching, Germany}
\affiliation{Munich Center for Quantum Science and Technology (MCQST), D-80799 Munich, Germany}
\affiliation{Blackett Laboratory, Imperial College London, London SW7 2AZ, United Kingdom}
\author{Roderich Moessner}
\affiliation{Max-Planck-Institut für Physik komplexer Systeme, Nöthnitzer Straße 38, D-01187 Dresden, Germany}
\author{N. P. Armitage}
\affiliation{Institute for Quantum Matter and Department of Physics and Astronomy, Johns Hopkins University, Baltimore MD 21218, USA}
\author{Tomohiro Takayama}
\affiliation{Max Planck Institute for Solid State Research, Heisenbergstrasse 1, 70569 Stuttgart, Germany}
\author{Hidenori Takagi}
\affiliation{Max Planck Institute for Solid State Research, Heisenbergstrasse 1, 70569 Stuttgart, Germany}
\affiliation{Department of Physics, The University of Tokyo, Bunkyo, 113-8654, Japan}
\author{Collin Broholm}
\affiliation{Institute for Quantum Matter and Department of Physics and Astronomy, Johns Hopkins University, Baltimore MD 21218, USA}
\affiliation{Department of Materials Science and Engineering,
The\ Johns\ Hopkins\ University, Baltimore, Maryland\ 21218, USA}
\affiliation{NIST Center for Neutron Research, Gaithersburg, Maryland\ 20899, USA}

\date{\today}

\begin{abstract}
Inelastic neutron scattering (INS) measurements of powder D$_3(^{7}$Li)($^{193}$Ir)$_2$O$_6$ reveal low energy magnetic excitations with a scattering cross section that is broad in $|Q|$ and consistent with a Kitaev spin-liquid (KSL) state. The magnetic nature of the excitation spectrum is demonstrated by longitudinally polarized neutron studies. The total magnetic moment of 1.7(2)$\mu_B$/Ir inferred from the total magnetic scattering cross section is consistent with the effective moment inferred from magnetic susceptibility data and expectations for the $J_{\rm eff}=1/2$ single ion state. The rise in the dynamic correlation function ${\cal S}(Q,\omega)$ for $\hbar\omega<5~$meV can be described by a nearest-neighbor Kitaev model with interaction strength $K\approx-13(5)$~meV. Exchange disorder associated with the mixed D-Li site could play an important role in stabilizing the low $T$ quantum fluctuating state. 
\end{abstract}

\maketitle
The exactly solvable Kitaev model \cite{Kitaev2006AnyonsBeyond} has catalyzed a surge of experimental effort to realize a Kitaev-like spin liquid (KSL). Based on $S$=1/2 spins on a honeycomb lattice with bond-dependent Ising exchange, the KSL features emergent anyonic $Z_2$ gauge fluxes and Majorana fermion excitations. Based on the work of Jackeli and Khaliullin~\cite{Jackeli2009MottModels}, several candidate magnetic materials with strong spin-orbit interactions were identified wherein Kitaev interactions play a significant role. These include A$_2$IrO$_3$ (A=Li,Na) \cite{Williams2016Incommensurate-Li2IrO3, Choi2019Spin-Li2IrO3, ChunNa2IrO3, Takayama2015HyperhoneycombMagnetism, Modic2014RealizationIridate} and $\alpha$-RuCl$_3$\cite{Plumb2014-Lattice, Banerjee2017Neutron-RuCl3, Banerjee2018Excitations-RuCl3, Sears2017}. However, these and most other KSL candidate materials develop long-range magnetic order at low $T$ apparently due to the presence of non-Kitaev interactions that are allowed by symmetry~\cite{Takagi2019concept}. The KSL can survive the presence of Heisenberg, Ising, and bond-dependent off-diagonal exchange that must be present in real materials, but only in a narrow window of parameter space\cite{Kimchi2011Kitaev-Heisenberg-J3, Rau2014GenericLimit, Katukuri2014KitaevCalculations, Gotfryd2017PhaseEffects, Chaloupka2010Kitaev-heisenbergA2IrO3}, making its materialization challenging. 

H$_3$LiIr$_2$O$_6$ is a rare example of a putative Kitaev material without long range magnetic order. In contrast to the X-Li$_2$IrO$_3$ (X=$\alpha,\beta,\gamma$) family of compounds, where the Kitaev interaction favors a three-dimensional long range ordered non-collinear spin structures~\cite{Kimchi2015UnifiedLi2IrO3, Biffin2014Noncoplanar-li2iro3, Modic2014RealizationIridate, Freund2016SingleOxides, Williams2016Incommensurate-Li2IrO3, HalloranLi2IrO3}, the interlayer Li$^+$ ions in $\alpha-$Li$_2$IrO$_3$ are replaced by H$^+$~\cite{Bette2017SolutionH3LiIr2O6} in H$_3$LiIr$_2$O$_6$. Thus the LiIr$_2$O$_6$ honeycomb plane is preserved but with reduced and disordered inter-layer exchange interactions due to positional disorder of the H$^+$ ions. Despite a Curie-Weiss temperature of $\Theta_{CW}=-105$~K~\cite{Kitagawa2018ALattice}, H$_3$LiIr$_2$O$_6$ has no magnetic phase transition down to temperatures as low as $T=50$~mK, and zero-field specific heat capacity data $C(T)/T\propto T^{-1/2}$ indicate a large density of low energy excited states.  NMR evidence that these low energy excitations are  magnetic~\cite{Kitagawa2018ALattice} is, however, inconsistent with a pure KSL~\cite{Kitagawa2018ALattice} and quantum paraelectric behavior~\cite{Geirhos2020QuantumO6, Wang2018PossibleQP} indicates that H$^+$ induced exchange disorder may play an important role in suppressing magnetic order as in related iridates~\cite{Yadav2018HLIOdisorder, Li2018RoleO6}. H$^+$ might promote a bond-disordered spin liquid~\cite{Knolle2019Bond-DisorderedHopping} or, under a specific stacking-fault pattern, a gapless spin liquid~\cite{Slagle2018TheoryO6}. A random singlet state might be able to account for the low-lying magnetic excitation, where $J_{\rm eff}=1/2$ moments form singlets over a  distribution of length scales prescribed by the quenched disorder~\cite{Kimchi2018ScalingSystems, LeeChanhyeonRandomSingletHLIO2023}.
\begin{figure}[t]
    \centering
    \includegraphics[]{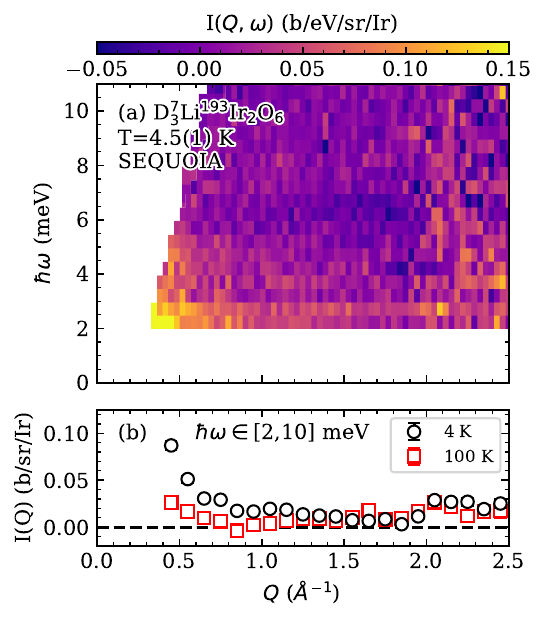}
    \caption{Magnetic spectrum of D$_3$LiIr$_2$O$_6$ measured on SEQUOIA after subtracting the non-magnetic background. (a) Normalized scattering intensity $I(Q,\omega)$ of D$_3$LiIr$_2$O$_6$ measured at $T=4.5(1)$~K. Spectrum within the elastic line of the experimental condition, i.e. $\hbar\omega<1.8$~ meV, is not presented. (b) Comparison of integrated scattering intensity $I(Q)=\int I(Q,\omega)d\omega$ in the range of $\hbar\omega=2-10$~meV at $T=4.5(1)$~K (black) and $T=100.0(1)$~K (red).}
    \label{fig:seq_mainfig}
\end{figure}
At much higher energies, Raman spectroscopy shows a dome-shaped continuum of magnetic excitations with maximum intensity at 33 meV ~\cite{Pei2020MagneticO6}. These data are consistent with the anticipated two-spinon process in a KSL. A recent resonant inelastic x-ray scattering (RIXS) measurement documented a temperature-dependent, momentum-independent excitation continuum with a  spectral weight maximum near 25 meV~\cite{delatorre2023momentumindependent}  consistent with Raman spectroscopy, which, along  with recent $\mu$SR measurements~\cite{yang2022muon}, were interpreted as evidence for a bond-disordered KSL in H$_3$LiIr$_2$O$_6$. Raman and non spin flip RIXS are sensitive only to pairs of Majorana excitations, meaning that the energy scale is shifted to higher energy scales than the low energy bare Majorana band structure, which has yet to be measured~\cite{knolle2014raman,Knolle2014DynamicsFluxes}.

In this Letter, we examine the scalar momentum $|Q|-$resolved spectrum of magnetic excitations in a D$_3$LiIr$_2$O$_6$ powder sample in the  $\hbar\omega<10$~meV energy range using  inelastic magnetic neutron scattering from a powder sample. We document magnetic spectra that account for most of the spectral weight of the $J_{\rm eff} = 1/2$ states of Ir$^{4+}$ with no dispersion apparent in the powder averaged continuum. The equal time correlation function can be described by nearest-neighbor correlations only and the broad spectral maximum at $\hbar\omega = 2.5(5)$ meV is consistent with Kitaev interactions $K\approx-13(5)$ meV~\cite{Knolle2014DynamicsFluxes, Nasu2021, Knolle2019Bond-DisorderedHopping, Samarakoon2017}. 

 \begin{figure}[t]
    \centering
    \includegraphics[]{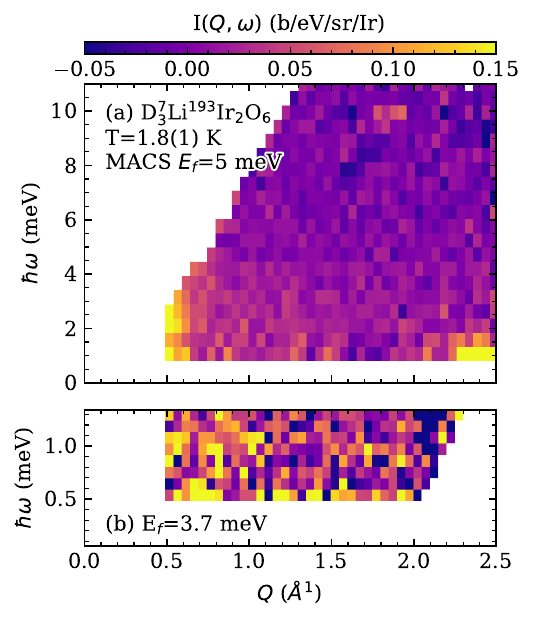}
    \caption{Magnetic inelastic scattering at $T=1.8(1)$~K obtained on MACS, using the E$_f$=5.0 meV (a) and E$_f$ = 3.7 meV (b) configurations. The colorbar scale is consistent with Fig.~\ref{fig:seq_mainfig}(a)}
    \label{fig:macs_data}
\end{figure}

\begin{figure}
    \centering
    \includegraphics[width=1.0\columnwidth]{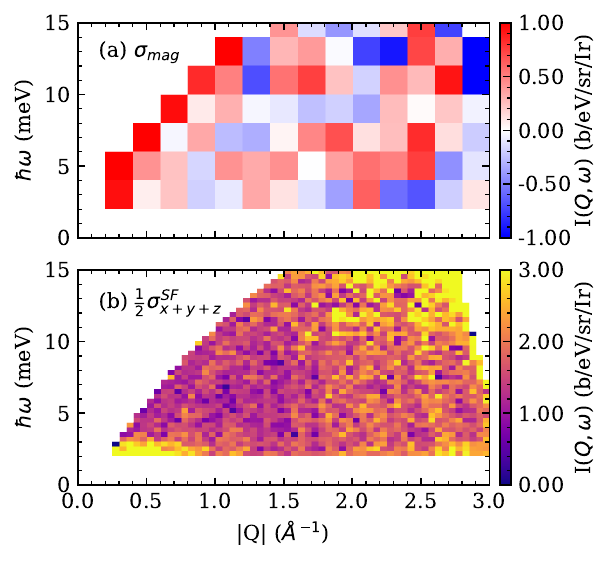}
    \caption{Polarized neutron scattering spectrum measured on HYSPEC. (a) Magnetic scattering cross-section $\sigma_{\rm mag}$ as defined in the texts. (b) Total spin-flip scattering cross-section.}
    \label{fig:HYS}
\end{figure}

The powder sample was prepared by previously published solid-state synthesis methods  \cite{Kitagawa2018ALattice} using $^2$H$=$D, $^{193}$Ir, and $^{7}$Li to mitigate absorption and incoherent scattering that is associated with the natural isotope distribution. The powder was held in an annular can with thickness $0.5$~mm and outer diameter $20$~mm resulting in $95$\% neutron transmission for 25 meV neutrons. Three inelastic neutron scattering experiments were performed. We used the SEQUOIA instrument at Oak Ridge National Laboratory (ORNL)~\cite{Stone2014ASource} to measure the spectrum down to $1.8$~meV for temperatures $T=4.0(1)$~K, $100.0(1)$~K, and $200.0(1)$~K (Fig.~\ref{fig:seq_mainfig} and Supplementary Information Fig. S1). Lower energy measurements were performed on the MACS spectrometer at the NIST Center for Neutron Research \cite{Rodriguez2008MACSNIST}, with sample temperatures $T=1.7(1)$~K and $T=55.0(1)$~K using the $E_f=3.7$~meV and $E_f=5$~meV configurations (Fig.~\ref{fig:macs_data}). Scattering angle and energy dependent absorption corrections were applied to all data. Nonmagnetic contributions to the scattering from H/D incoherent scattering and low-energy acoustic phonons were subtracted as described in the Supplementary Information. Absolute normalization of the scattering data was achieved by comparing the measured count rates to those for a vanadium standard  (SEQUOIA) and through the $Q-$integrated nuclear Bragg peak intensity (MACS). We used polarized neutrons on the HYSPEC instrument at ORNL \cite{Stone2014ASource} for an independent determination of the magnetic scattering cross section. Using three perpendicular guide field directions in succession, six cross-sections (the $x,y,z$, spin-flip and non-spin-flip channels) were measured at $T=2.0(1)$~K resulting in a model independent albeit low statistics measure of the magnetic scattering cross section (Fig.~\ref{fig:HYS}). Time domain Terahertz spectroscopy was performed on a custom-built spectrometer with a frequency range 0.2~THz to 2~THz \cite{Laurita2017LowMagnets} at zero magnetic field. Details of the data analysis provided in the supplementary information. 

We present the background corrected magnetic neutron scattering intensity measured on SEQUOIA in Fig. \ref{fig:seq_mainfig}.  While no sharp peaks in either $Q$ or $\hbar\omega$ were resolved down to the lowest accessible values of $Q=0.5~\AA^{-1}$ and $\hbar\omega=1.8$~meV as constrained by kinematics and contamination from elastic scattering, a buildup of intensity is apparent at  low-$Q$ and low-$E$. The wave vector dependence of the intensity $I(Q)$ integrated over $\hbar\omega\in [2,10]$~meV is shown in Fig.\ref{fig:seq_mainfig}(b). Upon cooling from 100~K to 4~K there is an increase in $I(Q)$ for  $Q<1.5~\AA^{-1}$. While contrary to any nuclear scattering cross section for a solid, this is consistent with expectations for an anisotropic quantum magnet dominated by dynamic spin correlations. 

Data from MACS covering $\hbar\omega\in[0.5,10]$~meV is in Fig.~\ref{fig:macs_data}(a,b). In the overlapping regimes of $(Q,\omega)$, the SEQUOIA (Fig.~\ref{fig:seq_mainfig}(a)) and MACS data (Fig.~\ref{fig:macs_data}(a,b)) are consistent, which provides an important consistency check on the methods used to isolate magnetic scattering for the different spectrometers and sample configurations. In the regime down to $\hbar\omega=0.5$~meV that is uniquely revealed by MACS, there is a flattening of spectral weight for $\hbar\omega<2$~meV that will be examined in greater detail below.

\begin{figure}
    \centering
    \includegraphics[width=1.0\columnwidth]{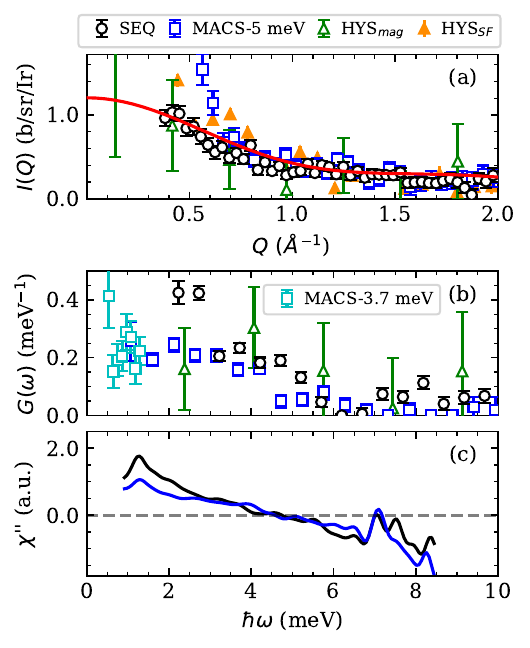}
    \caption{Energy- and momentum-dependence of magnetic spectrum. In panels (a) and (b) the blue points are extracted from the MACS E$_f$=5 meV measurement, cyan from the MACS E$_f$=3.7 meV measurement, black from the SEQUOIA measurement, green points are the $\sigma_{mag}$ contribution from the polarized HYSPEC experiment, and orange points are the $\sigma^{SF}_{x+y+z}/2 -a-bQ^2$ contribution from the HYSPEC experiment. (a) Momentum-dependence of the spectra ${\cal S}(Q)$ factorized from the neutron spectrum. The red line is a fit to the nearest-neighbor correlation function described in the text. (b) Energy-dependent spectra G$(\omega$) obtained from the same factorization analysis. Each individual factorization has been normalized following the practice described in the main texts. (c) THz spectroscopy taken at $T$=3 K for two samples.
    The temperature-dependence is presented in supplementary information, Fig. S16. All error bars represent one standard deviation.}
    \label{fig:factorizations}
\end{figure}

For a separate model independent determination of the magnetic scattering cross section in D$_3(^{7}$Li)($^{193}$Ir)$_2$O$_6$ we performed a fully polarized neutron scattering experiment on HYSPEC. For an isotropic sample such as a powder, the total magnetic scattering component is  given by $\sigma_{\rm mag} = 2(\sigma_{x}^{\rm SF}+\sigma_{y}^{\rm SF} - 2\sigma_{z}^{\rm SF})$~\cite{Scharpf1993TheXM}. Here $x,y,z$ label three perpendicular directions of the guide field for measurement of the spin flip (SF) and non-spin flip (NSF) part of the scattering cross section, $z$ being perpendicular to the scattering plane in our case. After proton charge of (150, 150, 330) Coulomb dedicated to measuring the $(\sigma_x^{\rm SF},\sigma_y^{\rm SF},\sigma_z^{\rm SF})$ cross sections respectively, the inferred data for $\sigma_{\rm mag}$ are shown in Fig.~\ref{fig:HYS}. Improved statistics for $\sigma_{\rm mag}$ can be obtained by averaging over $(Q,\omega)$ in coarse-grained cuts along $Q$ ($\hbar\omega$) direction while integrating over the entire range of $\hbar\omega$ ($Q$). Such data are presented in Figs.~\ref{fig:factorizations}(a and b) and they do show a magnetic contribution to the scattering cross section that is quantitatively consistent with the higher statistics unpolarized data. We also present  $(\sigma_{x}^{\rm SF}+\sigma_{y}^{\rm SF}+\sigma_{z}^{\rm SF})/2=\sigma_{\rm mag}+\frac{3}{2}\sigma_{\rm N}^{\rm inc}$ in Fig.~\ref{fig:HYS}(b)\cite{lovesey1984theory, Scharpf1993TheXM}. As the incoherent inelastic nuclear scattering cross section $\sigma_{\rm N}^{\rm inc}$ has no $Q-$dependence beyond the Debye-Waller factor, ${\rm exp}(-Q^2\langle u^2\rangle)Q^2$, and spin-incoherent phonon scattering $\propto Q^2$, the resemblance of the low-$Q$ part of $\sigma_{\rm mag}+\frac{3}{2}\sigma_{\rm N}^{\rm inc}$ (Fig.~\ref{fig:HYS}(b)) with $I(Q,\omega)$ presented in Figs.~\ref{fig:seq_mainfig} and~\ref{fig:macs_data} affirms the magnetic origin of the low-$Q$ scattering for $\hbar\omega<5$~meV and $Q<1$~\AA$^{-1}$. 

We now seek a quantitative comparison of the three different measurements of the magnetic scattering cross section. For improved statistical accuracy and to avoid systematic errors associated with the coverage of $Q-\omega$ space dictated by the kinematics of the scattering process, we project the data onto the $Q-$ and $\omega-$axes. For the unpolarized data we obtain values of $I(Q)$, of length $N_Q$, and $G(\omega)$, of length $N_\omega$, through a least-squared fit to the data under the assumption of a factorizable cross section: $I(Q,\omega)=I(Q)G(\omega)$. Here $G(\omega)$ is unity normalized $\int G(\omega)\hbar d\omega \equiv 1$ over the inclusive energy range $[0.5,10]$~meV covered by the overlapping data sets. This projects data from $\lesssim N_Q\times N_\omega$ pixels in $I(Q,\omega)$ to $N_Q+N_\omega$ pixels in $I(Q)$ and $G(\omega)$. Due to the lower counting statistics of the polarized data we found it more effective to directly integrate the data accommodating the kinematic limits as follows: $G(\omega)$ was obtained as the average of $I(Q,\omega)$ over $Q\in[0.2,1.6]~{\rm \AA}^{-1}$ scaled by a factor $f$ to enforce its unity normalization. $I(Q)$ was obtained as the average of $I(Q,\omega)$ over $\hbar\omega\in[2,10]~$meV scaled by a factor $1/f$ so that $I(Q,\omega)\approx I(Q)G(\omega)$. For a direct comparison of $I(Q)$ from $\frac{1}{2}\sigma_{x+y+z}^{SF}$ to the other measurements in Fig. \ref{fig:factorizations}(a), a background of the form $I_{bkg}(Q)=a+bQ^2$ was subtracted to account for nuclear spin-incoherent scattering. Superior statistics also allow for a finer binning than $\sigma_{mag}$ in Fig.~\ref{fig:factorizations}. 
Though acquired on different instruments, subjected to different background subtractions, and normalized to different reference cross sections, the data sets for $I(Q)$ and $G(\omega)$ displayed in Figs.~\ref{fig:factorizations}(a) and (b) provide statistically consistent measures of the magnetic scattering cross section in D$_3(^{7}$Li)($^{193}$Ir)$_2$O$_6$. 

From the total moment sum rule we have $\mu_{\rm eff}^2=(6/r_0^2)\int I(Q)/|F(Q)|^2 dQ$~\cite{lovesey1984theory}. Integrating over available data in the ranges $Q\in [0.5,1.7]~{\rm \AA}^{-1}$ and $\hbar\omega\in [1.8,10]$~meV yields $\mu_{\rm eff}=1.8(4)\ \mu_B$ for the unpolarized experiments and $\mu_{eff}=1.6(8)\ \mu_B$ for the polarized data, with error bars dominated by the uncertainty in normalization. These values are close to the effective moment inferred from high-temperature magnetic susceptibility data $\mu_{\rm eff}=1.60\ \mu_B$~\cite{Kitagawa2018ALattice} and consistent with $g\sqrt{J_{\rm eff}(J_{\rm eff}+1)}\mu_B$ with $g\approx 2$ and $J_{\rm eff}=1/2$ for Ir$^{4+}$. 

Inaccessible through neutron scattering, we obtain the $Q=0$ magnetic excitation spectrum  through time-domain THz spectroscopy. Fig.~\ref{fig:factorizations}(c) presents $\chi''(\omega)$ , which reveals low-energy excitation with energy scales similar to neutron scattering spectra. (Fig. S16 in supplementary information reports the $T-$dependence of $\chi''$). With excellent consistency across multiple distinct experiments, our data demonstrate a buildup of magnetic excitations for $Q<1.5~\AA^{-1}$ and $\hbar\omega<5$~meV, with maximal spectral weight near $2$~meV. The significant spectral weight near $Q=0$ consistently documented by neutron scattering and THz spectroscopy in the absence of an applied field indicates anisotropic ferromagnetic interactions.

The presence of a single broad maximum in $I(Q)$ at $Q=0$ indicates the spin correlations are dominated by a nearest neighbor length scale. In the pure Kitaev spin liquid, spin correlations beyond nearest neighbors vanish,  approximating the dynamic correlation function as ${\cal S}(\textbf{Q},\omega)=2({\cal S}_0(\omega)-\text{sgn}(K) \frac{1}{3}\sum_i\cos(\textbf{Q}\cdot \textbf{d}_i){\cal S}_1(\omega))$, where $\textbf{d}_i$ are the three vectors separating nearest neighbors~\cite{Knolle2014DynamicsFluxes} and $\text{sgn}(K)$ is the sign of the Kitaev interaction.

We find that the powder averaged form ${\cal S}(Q,\omega)=2({\cal S}_0(\omega)-\text{sgn}(K)  {\cal S}_1(\omega)\sin (Qd)/(Qd))$ provides an excellent account of $I(Q)$ (solid line in Fig. \ref{fig:factorizations}(a)) with $K<0$ corresponding to ferromagnetic Kitaev interactions. 
Consistent with this, studies of  $(\alpha,\beta)$-Li$_2$IrO$_3$, which has similar super-exchange paths to D$_3$LiIr$_2$O$_6$, indicate the nearest-neighbor Kitaev interaction is ferromagnetic even though the Curie-Weiss temperature is negative, implying further antialigning interactions, such as possible antiferromagnetic Heisenberg interactions~\cite{Singh2012Relevance3, Choi2012Spin3, Winter2016ChallengesScales, Kitagawa2018ALattice}. The ideal KSL has a gapful  excitation continuum with spectral weight peaked at $|K|/5$ \cite{Knolle2014DynamicsFluxes}. Although this value is subject to revision due to the presence of disorder in the current system \cite{Knolle2019Bond-DisorderedHopping}, we estimate $K=-13(5)$ meV from the broad maximum in $G(\omega)$ near $\hbar\omega=2.5(5)$~meV (Fig.~\ref{fig:factorizations}(b)). 

It is interesting to compare the present results to our recent neutron scattering experiments on $\beta-{\rm Li_2IrO_3}$\cite{HalloranLi2IrO3}. This material forms a 3D hyper-Honeycomb lattice with similar coordination and super-exchange connectivity for Ir$^{4+}$ ions. However, it develops incommensurate long range magnetic order for $T_{\rm N}=38~$K and in this ordered state the inelastic neutron scattering spectrum is radically different from that of D$_3$LiIr$_2$O$_6$. Although the experiments were both conducted on isotopically enriched powder samples under very similar experimental conditions, $\beta-{\rm Li_2IrO_3}$ develops a well-defined peak in the magnetic excitation spectrum at 12 meV with a 2.1 meV gap to excitations. There is also pronounced $Q-$dependence with low energy magnetic scattering onsetting near the incommensurate magnetic wave vector. While diffuse scattering associated with deuterium and D/Li site mixing complicates the present experiment, the comparison to $\beta-{\rm Li_2IrO_3}$ makes it clear that we have sensitivity to detect coherent spin wave like excitations and that magnetic excitations in D$_3$LiIr$_2$O$_6$ take a very different form. Spin wave theory using the $JK\Gamma$ model does provide a good account of the data for $\beta-{\rm Li_2IrO_3}$ with a Heisenberg term $J=0.40(2)~$meV, Kitaev term $K=-24(3)~$meV, and off-diagonal term $\Gamma=-9.3(1)~$meV~\cite{HalloranLi2IrO3}. Given the similar local coordination environment for the edge sharing Ir$^{4+}$ nearest neighbors we can expect similar exchange parameters for D$_3$LiIr$_2$O$_6$.

We have shown that the magnetic neutron scattering cross section for a powder sample of D$_3$LiIr$_2$O$_6$ is consistent with a Kitaev spin-liquid where the flux excitation gap is closed by disorder and interactions beyond the pure model. The increased spacing between Kitaev layers and the disordered nature of D/Li in the intervening layer leads to weak and disordered inter-layer interactions that surely play a role in suppressing long range order and thereby relatively favoring a dynamic spin-liquid-like state.  However,  compounded by powder averaging, the featureless nature of the magnetic scattering anticipated precludes distinguishing between the various scenarios proposed for this material, such as a pure KSL phase, a random-singlet state, and a bond-disordered spin-liquid phase~\cite{Slagle2018TheoryO6, Kao2020Vacancy-inducedLiquid,kao2023vacancy,Knolle2019Bond-DisorderedHopping}. Inelastic neutron scattering on single crystals will be needed for this purpose, but their availability is still limited~\cite{Pei2020MagneticO6, Freund2016SingleOxides}. 


\section{Acknowledgements}
This work was supported as part of the Institute for Quantum Matter, an Energy Frontier Research Center funded by the U.S. Department of Energy, Office of Science, Basic Energy Sciences under Award No. DE-SC0019331. CB was supported by the Gordon and Betty Moore foundation EPIQS program under GBMF9456. Access to MACS was provided by the Center for High Resolution Neutron Scattering, a partnership between the National Institute of Standards and Technology and the National Science Foundation under Agreement No. DMR-2010792. A portion of this research used resources at the Spallation Neutron Source, a DOE Office of Science User Facility operated by the Oak Ridge National Laboratory.  Research in Dresden was in part supported by the Deutsche
Forschungsgemeinschaft, under Grants No. SFB 1143
(Project No. 247310070) and the cluster of excellence
ct.qmat (EXC 2147, Project No. 390858490). J.K. acknowledges support from the Deutsche Forschungsgemeinschaft (DFG, German Research Foundation) under Germany’s Excellence Strategy– EXC–2111–390814868 and  TRR 360 - 492547816, as well as the Munich Quantum Valley, which is supported by the Bavarian state government with funds from the Hightech Agenda Bayern Plus. J.K. also acknowledges support from the Imperial-TUM flagship partnership. Identification of
commercial equipment does not imply recommendation or endorsement by NIST.

\bibliography{refs}

\begin{thebibliography}{48}%
\makeatletter
\providecommand \@ifxundefined [1]{%
 \@ifx{#1\undefined}
}%
\providecommand \@ifnum [1]{%
 \ifnum #1\expandafter \@firstoftwo
 \else \expandafter \@secondoftwo
 \fi
}%
\providecommand \@ifx [1]{%
 \ifx #1\expandafter \@firstoftwo
 \else \expandafter \@secondoftwo
 \fi
}%
\providecommand \natexlab [1]{#1}%
\providecommand \enquote  [1]{``#1''}%
\providecommand \bibnamefont  [1]{#1}%
\providecommand \bibfnamefont [1]{#1}%
\providecommand \citenamefont [1]{#1}%
\providecommand \href@noop [0]{\@secondoftwo}%
\providecommand \href [0]{\begingroup \@sanitize@url \@href}%
\providecommand \@href[1]{\@@startlink{#1}\@@href}%
\providecommand \@@href[1]{\endgroup#1\@@endlink}%
\providecommand \@sanitize@url [0]{\catcode `\\12\catcode `\$12\catcode `\&12\catcode `\#12\catcode `\^12\catcode `\_12\catcode `\%12\relax}%
\providecommand \@@startlink[1]{}%
\providecommand \@@endlink[0]{}%
\providecommand \url  [0]{\begingroup\@sanitize@url \@url }%
\providecommand \@url [1]{\endgroup\@href {#1}{\urlprefix }}%
\providecommand \urlprefix  [0]{URL }%
\providecommand \Eprint [0]{\href }%
\providecommand \doibase [0]{https://doi.org/}%
\providecommand \selectlanguage [0]{\@gobble}%
\providecommand \bibinfo  [0]{\@secondoftwo}%
\providecommand \bibfield  [0]{\@secondoftwo}%
\providecommand \translation [1]{[#1]}%
\providecommand \BibitemOpen [0]{}%
\providecommand \bibitemStop [0]{}%
\providecommand \bibitemNoStop [0]{.\EOS\space}%
\providecommand \EOS [0]{\spacefactor3000\relax}%
\providecommand \BibitemShut  [1]{\csname bibitem#1\endcsname}%
\let\auto@bib@innerbib\@empty
\bibitem [{\citenamefont {Kitaev}(2006)}]{Kitaev2006AnyonsBeyond}%
  \BibitemOpen
  \bibfield  {author} {\bibinfo {author} {\bibfnamefont {A.}~\bibnamefont {Kitaev}},\ }\bibfield  {title} {\bibinfo {title} {{Anyons in an exactly solved model and beyond}},\ }\href {https://doi.org/10.1016/j.aop.2005.10.005} {\bibfield  {journal} {\bibinfo  {journal} {Annals of Physics}\ }\textbf {\bibinfo {volume} {321}},\ \bibinfo {pages} {2} (\bibinfo {year} {2006})}\BibitemShut {NoStop}%
\bibitem [{\citenamefont {Jackeli}\ and\ \citenamefont {Khaliullin}(2009)}]{Jackeli2009MottModels}%
  \BibitemOpen
  \bibfield  {author} {\bibinfo {author} {\bibfnamefont {G.}~\bibnamefont {Jackeli}}\ and\ \bibinfo {author} {\bibfnamefont {G.}~\bibnamefont {Khaliullin}},\ }\bibfield  {title} {\bibinfo {title} {{Mott insulators in the strong spin-orbit coupling Limit: From Heisenberg to a Quantum Compass and Kitaev Models}},\ }\bibfield  {journal} {\bibinfo  {journal} {Physical Review Letters}\ }\textbf {\bibinfo {volume} {102}},\ \href {https://doi.org/10.1103/PhysRevLett.102.017205} {10.1103/PhysRevLett.102.017205} (\bibinfo {year} {2009})\BibitemShut {NoStop}%
\bibitem [{\citenamefont {Williams}\ \emph {et~al.}(2016)\citenamefont {Williams}, \citenamefont {Johnson}, \citenamefont {Freund}, \citenamefont {Choi}, \citenamefont {Jesche}, \citenamefont {Kimchi}, \citenamefont {Manni}, \citenamefont {Bombardi}, \citenamefont {Manuel}, \citenamefont {Gegenwart},\ and\ \citenamefont {Coldea}}]{Williams2016Incommensurate-Li2IrO3}%
  \BibitemOpen
  \bibfield  {author} {\bibinfo {author} {\bibfnamefont {S.~C.}\ \bibnamefont {Williams}}, \bibinfo {author} {\bibfnamefont {R.~D.}\ \bibnamefont {Johnson}}, \bibinfo {author} {\bibfnamefont {F.}~\bibnamefont {Freund}}, \bibinfo {author} {\bibfnamefont {S.}~\bibnamefont {Choi}}, \bibinfo {author} {\bibfnamefont {A.}~\bibnamefont {Jesche}}, \bibinfo {author} {\bibfnamefont {I.}~\bibnamefont {Kimchi}}, \bibinfo {author} {\bibfnamefont {S.}~\bibnamefont {Manni}}, \bibinfo {author} {\bibfnamefont {A.}~\bibnamefont {Bombardi}}, \bibinfo {author} {\bibfnamefont {P.}~\bibnamefont {Manuel}}, \bibinfo {author} {\bibfnamefont {P.}~\bibnamefont {Gegenwart}},\ and\ \bibinfo {author} {\bibfnamefont {R.}~\bibnamefont {Coldea}},\ }\bibfield  {title} {\bibinfo {title} {{Incommensurate counterrotating magnetic order stabilized by Kitaev interactions in the layered honeycomb {$\alpha$}-Li$_2$IrO$_3$}},\ }\href {https://doi.org/10.1103/PhysRevB.93.195158} {\bibfield  {journal} {\bibinfo  {journal} {Physical Review B}\ }\textbf
  {\bibinfo {volume} {93}},\ \bibinfo {pages} {195158} (\bibinfo {year} {2016})}\BibitemShut {NoStop}%
\bibitem [{\citenamefont {Choi}\ \emph {et~al.}(2019)\citenamefont {Choi}, \citenamefont {Manni}, \citenamefont {Singleton}, \citenamefont {Topping}, \citenamefont {Lancaster}, \citenamefont {Blundell}, \citenamefont {Adroja}, \citenamefont {Zapf}, \citenamefont {Gegenwart},\ and\ \citenamefont {Coldea}}]{Choi2019Spin-Li2IrO3}%
  \BibitemOpen
  \bibfield  {author} {\bibinfo {author} {\bibfnamefont {S.}~\bibnamefont {Choi}}, \bibinfo {author} {\bibfnamefont {S.}~\bibnamefont {Manni}}, \bibinfo {author} {\bibfnamefont {J.}~\bibnamefont {Singleton}}, \bibinfo {author} {\bibfnamefont {C.~V.}\ \bibnamefont {Topping}}, \bibinfo {author} {\bibfnamefont {T.}~\bibnamefont {Lancaster}}, \bibinfo {author} {\bibfnamefont {S.~J.}\ \bibnamefont {Blundell}}, \bibinfo {author} {\bibfnamefont {D.~T.}\ \bibnamefont {Adroja}}, \bibinfo {author} {\bibfnamefont {V.}~\bibnamefont {Zapf}}, \bibinfo {author} {\bibfnamefont {P.}~\bibnamefont {Gegenwart}},\ and\ \bibinfo {author} {\bibfnamefont {R.}~\bibnamefont {Coldea}},\ }\bibfield  {title} {\bibinfo {title} {{Spin dynamics and field-induced magnetic phase transition in the honeycomb Kitaev magnet {$\alpha$}-Li$_2$IrO$_3$}},\ }\bibfield  {journal} {\bibinfo  {journal} {Physical Review B}\ }\textbf {\bibinfo {volume} {99}},\ \href {https://doi.org/10.1103/PhysRevB.99.054426} {10.1103/PhysRevB.99.054426} (\bibinfo {year}
  {2019})\BibitemShut {NoStop}%
\bibitem [{\citenamefont {{Hwan Chun}}\ \emph {et~al.}(2015)\citenamefont {{Hwan Chun}}, \citenamefont {{Kim}}, \citenamefont {{Kim}}, \citenamefont {{Zheng}}, \citenamefont {{Stoumpos}}, \citenamefont {{Malliakas}}, \citenamefont {{Mitchell}}, \citenamefont {{Mehlawat}}, \citenamefont {{Singh}}, \citenamefont {{Choi}}, \citenamefont {{Gog}}, \citenamefont {{Al-Zein}}, \citenamefont {{Sala}}, \citenamefont {{Krisch}}, \citenamefont {{Chaloupka}}, \citenamefont {{Jackeli}}, \citenamefont {{Khaliullin}},\ and\ \citenamefont {{Kim}}}]{ChunNa2IrO3}%
  \BibitemOpen
  \bibfield  {author} {\bibinfo {author} {\bibfnamefont {S.}~\bibnamefont {{Hwan Chun}}}, \bibinfo {author} {\bibfnamefont {J.-W.}\ \bibnamefont {{Kim}}}, \bibinfo {author} {\bibfnamefont {J.}~\bibnamefont {{Kim}}}, \bibinfo {author} {\bibfnamefont {H.}~\bibnamefont {{Zheng}}}, \bibinfo {author} {\bibfnamefont {C.~C.}\ \bibnamefont {{Stoumpos}}}, \bibinfo {author} {\bibfnamefont {C.~D.}\ \bibnamefont {{Malliakas}}}, \bibinfo {author} {\bibfnamefont {J.~F.}\ \bibnamefont {{Mitchell}}}, \bibinfo {author} {\bibfnamefont {K.}~\bibnamefont {{Mehlawat}}}, \bibinfo {author} {\bibfnamefont {Y.}~\bibnamefont {{Singh}}}, \bibinfo {author} {\bibfnamefont {Y.}~\bibnamefont {{Choi}}}, \bibinfo {author} {\bibfnamefont {T.}~\bibnamefont {{Gog}}}, \bibinfo {author} {\bibfnamefont {A.}~\bibnamefont {{Al-Zein}}}, \bibinfo {author} {\bibfnamefont {M.~M.}\ \bibnamefont {{Sala}}}, \bibinfo {author} {\bibfnamefont {M.}~\bibnamefont {{Krisch}}}, \bibinfo {author} {\bibfnamefont {J.}~\bibnamefont {{Chaloupka}}}, \bibinfo {author}
  {\bibfnamefont {G.}~\bibnamefont {{Jackeli}}}, \bibinfo {author} {\bibfnamefont {G.}~\bibnamefont {{Khaliullin}}},\ and\ \bibinfo {author} {\bibfnamefont {B.~J.}\ \bibnamefont {{Kim}}},\ }\bibfield  {title} {\bibinfo {title} {{Direct evidence for dominant bond-directional interactions in a honeycomb lattice iridate Na$_{2}$IrO$_{3}$}},\ }\href {https://doi.org/10.1038/nphys3322} {\bibfield  {journal} {\bibinfo  {journal} {Nature Physics}\ }\textbf {\bibinfo {volume} {11}},\ \bibinfo {pages} {462} (\bibinfo {year} {2015})}\BibitemShut {NoStop}%
\bibitem [{\citenamefont {Takayama}\ \emph {et~al.}(2015)\citenamefont {Takayama}, \citenamefont {Kato}, \citenamefont {Dinnebier}, \citenamefont {Nuss}, \citenamefont {Kono}, \citenamefont {Veiga}, \citenamefont {Fabbris}, \citenamefont {Haskel},\ and\ \citenamefont {Takagi}}]{Takayama2015HyperhoneycombMagnetism}%
  \BibitemOpen
  \bibfield  {author} {\bibinfo {author} {\bibfnamefont {T.}~\bibnamefont {Takayama}}, \bibinfo {author} {\bibfnamefont {A.}~\bibnamefont {Kato}}, \bibinfo {author} {\bibfnamefont {R.}~\bibnamefont {Dinnebier}}, \bibinfo {author} {\bibfnamefont {J.}~\bibnamefont {Nuss}}, \bibinfo {author} {\bibfnamefont {H.}~\bibnamefont {Kono}}, \bibinfo {author} {\bibfnamefont {L.~S.}\ \bibnamefont {Veiga}}, \bibinfo {author} {\bibfnamefont {G.}~\bibnamefont {Fabbris}}, \bibinfo {author} {\bibfnamefont {D.}~\bibnamefont {Haskel}},\ and\ \bibinfo {author} {\bibfnamefont {H.}~\bibnamefont {Takagi}},\ }\bibfield  {title} {\bibinfo {title} {{Hyperhoneycomb iridate {$\beta$}-Li$_2$IrO$_3$ as a platform for kitaev magnetism}},\ }\bibfield  {journal} {\bibinfo  {journal} {Physical Review Letters}\ }\textbf {\bibinfo {volume} {114}},\ \href {https://doi.org/10.1103/PhysRevLett.114.077202} {10.1103/PhysRevLett.114.077202} (\bibinfo {year} {2015})\BibitemShut {NoStop}%
\bibitem [{\citenamefont {Modic}\ \emph {et~al.}(2014)\citenamefont {Modic}, \citenamefont {Smidt}, \citenamefont {Kimchi}, \citenamefont {Breznay}, \citenamefont {Biffin}, \citenamefont {Choi}, \citenamefont {Johnson}, \citenamefont {Coldea}, \citenamefont {Watkins-Curry}, \citenamefont {McCandless}, \citenamefont {Chan}, \citenamefont {Gandara}, \citenamefont {Islam}, \citenamefont {Vishwanath}, \citenamefont {Shekhter}, \citenamefont {McDonald},\ and\ \citenamefont {Analytis}}]{Modic2014RealizationIridate}%
  \BibitemOpen
  \bibfield  {author} {\bibinfo {author} {\bibfnamefont {K.~A.}\ \bibnamefont {Modic}}, \bibinfo {author} {\bibfnamefont {T.~E.}\ \bibnamefont {Smidt}}, \bibinfo {author} {\bibfnamefont {I.}~\bibnamefont {Kimchi}}, \bibinfo {author} {\bibfnamefont {N.~P.}\ \bibnamefont {Breznay}}, \bibinfo {author} {\bibfnamefont {A.}~\bibnamefont {Biffin}}, \bibinfo {author} {\bibfnamefont {S.}~\bibnamefont {Choi}}, \bibinfo {author} {\bibfnamefont {R.~D.}\ \bibnamefont {Johnson}}, \bibinfo {author} {\bibfnamefont {R.}~\bibnamefont {Coldea}}, \bibinfo {author} {\bibfnamefont {P.}~\bibnamefont {Watkins-Curry}}, \bibinfo {author} {\bibfnamefont {G.~T.}\ \bibnamefont {McCandless}}, \bibinfo {author} {\bibfnamefont {J.~Y.}\ \bibnamefont {Chan}}, \bibinfo {author} {\bibfnamefont {F.}~\bibnamefont {Gandara}}, \bibinfo {author} {\bibfnamefont {Z.}~\bibnamefont {Islam}}, \bibinfo {author} {\bibfnamefont {A.}~\bibnamefont {Vishwanath}}, \bibinfo {author} {\bibfnamefont {A.}~\bibnamefont {Shekhter}}, \bibinfo {author} {\bibfnamefont
  {R.~D.}\ \bibnamefont {McDonald}},\ and\ \bibinfo {author} {\bibfnamefont {J.~G.}\ \bibnamefont {Analytis}},\ }\bibfield  {title} {\bibinfo {title} {{Realization of a three-dimensional spin-anisotropic harmonic honeycomb iridate}},\ }\bibfield  {journal} {\bibinfo  {journal} {Nature Communications}\ }\textbf {\bibinfo {volume} {5}},\ \href {https://doi.org/10.1038/ncomms5203} {10.1038/ncomms5203} (\bibinfo {year} {2014})\BibitemShut {NoStop}%
\bibitem [{\citenamefont {Plumb}\ \emph {et~al.}(2014)\citenamefont {Plumb}, \citenamefont {Clancy}, \citenamefont {Sandilands}, \citenamefont {Shankar}, \citenamefont {Hu}, \citenamefont {Burch}, \citenamefont {Kee},\ and\ \citenamefont {Kim}}]{Plumb2014-Lattice}%
  \BibitemOpen
  \bibfield  {author} {\bibinfo {author} {\bibfnamefont {K.~W.}\ \bibnamefont {Plumb}}, \bibinfo {author} {\bibfnamefont {J.~P.}\ \bibnamefont {Clancy}}, \bibinfo {author} {\bibfnamefont {L.~J.}\ \bibnamefont {Sandilands}}, \bibinfo {author} {\bibfnamefont {V.~V.}\ \bibnamefont {Shankar}}, \bibinfo {author} {\bibfnamefont {Y.~F.}\ \bibnamefont {Hu}}, \bibinfo {author} {\bibfnamefont {K.~S.}\ \bibnamefont {Burch}}, \bibinfo {author} {\bibfnamefont {H.~Y.}\ \bibnamefont {Kee}},\ and\ \bibinfo {author} {\bibfnamefont {Y.~J.}\ \bibnamefont {Kim}},\ }\bibfield  {title} {\bibinfo {title} {{{$\alpha$}-RuCl$_3$: A spin-orbit assisted Mott insulator on a honeycomb lattice}},\ }\bibfield  {journal} {\bibinfo  {journal} {Physical Review B - Condensed Matter and Materials Physics}\ }\textbf {\bibinfo {volume} {90}},\ \href {https://doi.org/10.1103/PhysRevB.90.041112} {10.1103/PhysRevB.90.041112} (\bibinfo {year} {2014})\BibitemShut {NoStop}%
\bibitem [{\citenamefont {Banerjee}\ \emph {et~al.}(2017)\citenamefont {Banerjee}, \citenamefont {Yan}, \citenamefont {Knolle}, \citenamefont {Bridges}, \citenamefont {Stone}, \citenamefont {Lumsden}, \citenamefont {Mandrus}, \citenamefont {Tennant}, \citenamefont {Moessner},\ and\ \citenamefont {Nagler}}]{Banerjee2017Neutron-RuCl3}%
  \BibitemOpen
  \bibfield  {author} {\bibinfo {author} {\bibfnamefont {A.}~\bibnamefont {Banerjee}}, \bibinfo {author} {\bibfnamefont {J.}~\bibnamefont {Yan}}, \bibinfo {author} {\bibfnamefont {J.}~\bibnamefont {Knolle}}, \bibinfo {author} {\bibfnamefont {C.~A.}\ \bibnamefont {Bridges}}, \bibinfo {author} {\bibfnamefont {M.~B.}\ \bibnamefont {Stone}}, \bibinfo {author} {\bibfnamefont {M.~D.}\ \bibnamefont {Lumsden}}, \bibinfo {author} {\bibfnamefont {D.~G.}\ \bibnamefont {Mandrus}}, \bibinfo {author} {\bibfnamefont {D.~A.}\ \bibnamefont {Tennant}}, \bibinfo {author} {\bibfnamefont {R.}~\bibnamefont {Moessner}},\ and\ \bibinfo {author} {\bibfnamefont {S.~E.}\ \bibnamefont {Nagler}},\ }\bibfield  {title} {\bibinfo {title} {{Neutron scattering in the proximate quantum spin liquid {$\alpha$}-RuCl$_3$}},\ }\href {https://doi.org/10.1126/science.aah6015} {\bibfield  {journal} {\bibinfo  {journal} {Science}\ }\textbf {\bibinfo {volume} {356}},\ \bibinfo {pages} {1055} (\bibinfo {year} {2017})}\BibitemShut {NoStop}%
\bibitem [{\citenamefont {Banerjee}\ \emph {et~al.}(2018)\citenamefont {Banerjee}, \citenamefont {Lampen-Kelley}, \citenamefont {Knolle}, \citenamefont {Balz}, \citenamefont {Aczel}, \citenamefont {Winn}, \citenamefont {Liu}, \citenamefont {Pajerowski}, \citenamefont {Yan}, \citenamefont {Bridges}, \citenamefont {Savici}, \citenamefont {Chakoumakos}, \citenamefont {Lumsden}, \citenamefont {Tennant}, \citenamefont {Moessner}, \citenamefont {Mandrus},\ and\ \citenamefont {Nagler}}]{Banerjee2018Excitations-RuCl3}%
  \BibitemOpen
  \bibfield  {author} {\bibinfo {author} {\bibfnamefont {A.}~\bibnamefont {Banerjee}}, \bibinfo {author} {\bibfnamefont {P.}~\bibnamefont {Lampen-Kelley}}, \bibinfo {author} {\bibfnamefont {J.}~\bibnamefont {Knolle}}, \bibinfo {author} {\bibfnamefont {C.}~\bibnamefont {Balz}}, \bibinfo {author} {\bibfnamefont {A.~A.}\ \bibnamefont {Aczel}}, \bibinfo {author} {\bibfnamefont {B.}~\bibnamefont {Winn}}, \bibinfo {author} {\bibfnamefont {Y.}~\bibnamefont {Liu}}, \bibinfo {author} {\bibfnamefont {D.}~\bibnamefont {Pajerowski}}, \bibinfo {author} {\bibfnamefont {J.}~\bibnamefont {Yan}}, \bibinfo {author} {\bibfnamefont {C.~A.}\ \bibnamefont {Bridges}}, \bibinfo {author} {\bibfnamefont {A.~T.}\ \bibnamefont {Savici}}, \bibinfo {author} {\bibfnamefont {B.~C.}\ \bibnamefont {Chakoumakos}}, \bibinfo {author} {\bibfnamefont {M.~D.}\ \bibnamefont {Lumsden}}, \bibinfo {author} {\bibfnamefont {D.~A.}\ \bibnamefont {Tennant}}, \bibinfo {author} {\bibfnamefont {R.}~\bibnamefont {Moessner}}, \bibinfo {author} {\bibfnamefont
  {D.~G.}\ \bibnamefont {Mandrus}},\ and\ \bibinfo {author} {\bibfnamefont {S.~E.}\ \bibnamefont {Nagler}},\ }\bibfield  {title} {\bibinfo {title} {{Excitations in the field-induced quantum spin liquid state of {$\alpha$}-RuCl$_3$}},\ }\bibfield  {journal} {\bibinfo  {journal} {npj Quantum Materials}\ }\textbf {\bibinfo {volume} {3}},\ \href {https://doi.org/10.1038/s41535-018-0079-2} {10.1038/s41535-018-0079-2} (\bibinfo {year} {2018})\BibitemShut {NoStop}%
\bibitem [{\citenamefont {Sears}\ \emph {et~al.}(2017)\citenamefont {Sears}, \citenamefont {Zhao}, \citenamefont {Xu}, \citenamefont {Lynn},\ and\ \citenamefont {Kim}}]{Sears2017}%
  \BibitemOpen
  \bibfield  {author} {\bibinfo {author} {\bibfnamefont {J.~A.}\ \bibnamefont {Sears}}, \bibinfo {author} {\bibfnamefont {Y.}~\bibnamefont {Zhao}}, \bibinfo {author} {\bibfnamefont {Z.}~\bibnamefont {Xu}}, \bibinfo {author} {\bibfnamefont {J.~W.}\ \bibnamefont {Lynn}},\ and\ \bibinfo {author} {\bibfnamefont {Y.~J.}\ \bibnamefont {Kim}},\ }\bibfield  {title} {\bibinfo {title} {Phase diagram of $\alpha$-rucl$_3$ in an in-plane magnetic field},\ }\href {https://doi.org/10.1103/PhysRevB.95.180411} {\bibfield  {journal} {\bibinfo  {journal} {Physical Review B}\ }\textbf {\bibinfo {volume} {95}},\ \bibinfo {pages} {180411} (\bibinfo {year} {2017})}\BibitemShut {NoStop}%
\bibitem [{\citenamefont {Takagi}\ \emph {et~al.}(2019)\citenamefont {Takagi}, \citenamefont {Takayama}, \citenamefont {Jackeli}, \citenamefont {Khaliullin},\ and\ \citenamefont {Nagler}}]{Takagi2019concept}%
  \BibitemOpen
  \bibfield  {author} {\bibinfo {author} {\bibfnamefont {H.}~\bibnamefont {Takagi}}, \bibinfo {author} {\bibfnamefont {T.}~\bibnamefont {Takayama}}, \bibinfo {author} {\bibfnamefont {G.}~\bibnamefont {Jackeli}}, \bibinfo {author} {\bibfnamefont {G.}~\bibnamefont {Khaliullin}},\ and\ \bibinfo {author} {\bibfnamefont {S.~E.}\ \bibnamefont {Nagler}},\ }\bibfield  {title} {\bibinfo {title} {Concept and realization of kitaev quantum spin liquids},\ }\href {https://doi.org/10.1038/s42254-019-0038-2} {\bibfield  {journal} {\bibinfo  {journal} {Nature Reviews Physics}\ }\textbf {\bibinfo {volume} {1}},\ \bibinfo {pages} {264} (\bibinfo {year} {2019})}\BibitemShut {NoStop}%
\bibitem [{\citenamefont {Kimchi}\ and\ \citenamefont {You}(2011)}]{Kimchi2011Kitaev-Heisenberg-J3}%
  \BibitemOpen
  \bibfield  {author} {\bibinfo {author} {\bibfnamefont {I.}~\bibnamefont {Kimchi}}\ and\ \bibinfo {author} {\bibfnamefont {Y.-Z.}\ \bibnamefont {You}},\ }\bibfield  {title} {\bibinfo {title} {{Kitaev-Heisenberg-$J_2$-$J_3$ model for the iridates A$_2$IrO$_3$}},\ }\href {https://doi.org/10.1103/PhysRevB.84.180407} {\bibfield  {journal} {\bibinfo  {journal} {RAPID COMMUNICATIONS PHYSICAL REVIEW B}\ }\textbf {\bibinfo {volume} {84}},\ \bibinfo {pages} {180407} (\bibinfo {year} {2011})}\BibitemShut {NoStop}%
\bibitem [{\citenamefont {Rau}\ \emph {et~al.}(2014)\citenamefont {Rau}, \citenamefont {Lee},\ and\ \citenamefont {Kee}}]{Rau2014GenericLimit}%
  \BibitemOpen
  \bibfield  {author} {\bibinfo {author} {\bibfnamefont {J.~G.}\ \bibnamefont {Rau}}, \bibinfo {author} {\bibfnamefont {E.~K.~H.}\ \bibnamefont {Lee}},\ and\ \bibinfo {author} {\bibfnamefont {H.~Y.}\ \bibnamefont {Kee}},\ }\bibfield  {title} {\bibinfo {title} {{Generic spin model for the honeycomb iridates beyond the Kitaev limit}},\ }\bibfield  {journal} {\bibinfo  {journal} {Physical Review Letters}\ }\textbf {\bibinfo {volume} {112}},\ \href {https://doi.org/10.1103/PhysRevLett.112.077204} {10.1103/PhysRevLett.112.077204} (\bibinfo {year} {2014})\BibitemShut {NoStop}%
\bibitem [{\citenamefont {Katukuri}\ \emph {et~al.}(2014)\citenamefont {Katukuri}, \citenamefont {Nishimoto}, \citenamefont {Yushankhai}, \citenamefont {Stoyanova}, \citenamefont {Kandpal}, \citenamefont {Choi}, \citenamefont {Coldea}, \citenamefont {Rousochatzakis}, \citenamefont {Hozoi},\ and\ \citenamefont {Brink}}]{Katukuri2014KitaevCalculations}%
  \BibitemOpen
  \bibfield  {author} {\bibinfo {author} {\bibfnamefont {V.~M.}\ \bibnamefont {Katukuri}}, \bibinfo {author} {\bibfnamefont {S.}~\bibnamefont {Nishimoto}}, \bibinfo {author} {\bibfnamefont {V.}~\bibnamefont {Yushankhai}}, \bibinfo {author} {\bibfnamefont {A.}~\bibnamefont {Stoyanova}}, \bibinfo {author} {\bibfnamefont {H.}~\bibnamefont {Kandpal}}, \bibinfo {author} {\bibfnamefont {S.}~\bibnamefont {Choi}}, \bibinfo {author} {\bibfnamefont {R.}~\bibnamefont {Coldea}}, \bibinfo {author} {\bibfnamefont {I.}~\bibnamefont {Rousochatzakis}}, \bibinfo {author} {\bibfnamefont {L.}~\bibnamefont {Hozoi}},\ and\ \bibinfo {author} {\bibfnamefont {J.~V.~D.}\ \bibnamefont {Brink}},\ }\bibfield  {title} {\bibinfo {title} {{Kitaev interactions between j = 1/2 moments in honeycomb Na$_2$IrO$_3$ are large and ferromagnetic: Insights from ab initio quantum chemistry calculations}},\ }\bibfield  {journal} {\bibinfo  {journal} {New Journal of Physics}\ }\textbf {\bibinfo {volume} {16}},\ \href
  {https://doi.org/10.1088/1367-2630/16/1/013056} {10.1088/1367-2630/16/1/013056} (\bibinfo {year} {2014})\BibitemShut {NoStop}%
\bibitem [{\citenamefont {Gotfryd}\ \emph {et~al.}(2017)\citenamefont {Gotfryd}, \citenamefont {Rusna{\v{c}}ko}, \citenamefont {Wohlfeld}, \citenamefont {Jackeli}, \citenamefont {Chaloupka},\ and\ \citenamefont {Ole{\'{s}}}}]{Gotfryd2017PhaseEffects}%
  \BibitemOpen
  \bibfield  {author} {\bibinfo {author} {\bibfnamefont {D.}~\bibnamefont {Gotfryd}}, \bibinfo {author} {\bibfnamefont {J.}~\bibnamefont {Rusna{\v{c}}ko}}, \bibinfo {author} {\bibfnamefont {K.}~\bibnamefont {Wohlfeld}}, \bibinfo {author} {\bibfnamefont {G.}~\bibnamefont {Jackeli}}, \bibinfo {author} {\bibfnamefont {J.}~\bibnamefont {Chaloupka}},\ and\ \bibinfo {author} {\bibfnamefont {A.~M.}\ \bibnamefont {Ole{\'{s}}}},\ }\bibfield  {title} {\bibinfo {title} {{Phase diagram and spin correlations of the Kitaev-Heisenberg model: Importance of quantum effects}},\ }\href {https://doi.org/10.1103/PhysRevB.95.024426} {\bibfield  {journal} {\bibinfo  {journal} {Physical Review B}\ }\textbf {\bibinfo {volume} {95}},\ \bibinfo {pages} {24426} (\bibinfo {year} {2017})}\BibitemShut {NoStop}%
\bibitem [{\citenamefont {Chaloupka}\ \emph {et~al.}(2010)\citenamefont {Chaloupka}, \citenamefont {Jackeli},\ and\ \citenamefont {Khaliullin}}]{Chaloupka2010Kitaev-heisenbergA2IrO3}%
  \BibitemOpen
  \bibfield  {author} {\bibinfo {author} {\bibfnamefont {J.}~\bibnamefont {Chaloupka}}, \bibinfo {author} {\bibfnamefont {G.}~\bibnamefont {Jackeli}},\ and\ \bibinfo {author} {\bibfnamefont {G.}~\bibnamefont {Khaliullin}},\ }\bibfield  {title} {\bibinfo {title} {{Kitaev-heisenberg model on a honeycomb lattice: Possible exotic phases in iridium oxides A$_2$IrO$_3$}},\ }\bibfield  {journal} {\bibinfo  {journal} {Physical Review Letters}\ }\textbf {\bibinfo {volume} {105}},\ \href {https://doi.org/10.1103/PhysRevLett.105.027204} {10.1103/PhysRevLett.105.027204} (\bibinfo {year} {2010})\BibitemShut {NoStop}%
\bibitem [{\citenamefont {Kimchi}\ \emph {et~al.}(2015)\citenamefont {Kimchi}, \citenamefont {Coldea},\ and\ \citenamefont {Vishwanath}}]{Kimchi2015UnifiedLi2IrO3}%
  \BibitemOpen
  \bibfield  {author} {\bibinfo {author} {\bibfnamefont {I.}~\bibnamefont {Kimchi}}, \bibinfo {author} {\bibfnamefont {R.}~\bibnamefont {Coldea}},\ and\ \bibinfo {author} {\bibfnamefont {A.}~\bibnamefont {Vishwanath}},\ }\bibfield  {title} {\bibinfo {title} {{Unified theory of spiral magnetism in the harmonic-honeycomb iridates {$\alpha$},{$\beta$}, and {$\gamma$} Li$_2$IrO$_3$}},\ }\href {https://doi.org/10.1103/PhysRevB.91.245134} {\bibfield  {journal} {\bibinfo  {journal} {Physical Review B - Condensed Matter and Materials Physics}\ }\textbf {\bibinfo {volume} {91}},\ \bibinfo {pages} {245134} (\bibinfo {year} {2015})}\BibitemShut {NoStop}%
\bibitem [{\citenamefont {Biffin}\ \emph {et~al.}(2014)\citenamefont {Biffin}, \citenamefont {Johnson}, \citenamefont {Kimchi}, \citenamefont {Morris}, \citenamefont {Bombardi}, \citenamefont {Analytis}, \citenamefont {Vishwanath},\ and\ \citenamefont {Coldea}}]{Biffin2014Noncoplanar-li2iro3}%
  \BibitemOpen
  \bibfield  {author} {\bibinfo {author} {\bibfnamefont {A.}~\bibnamefont {Biffin}}, \bibinfo {author} {\bibfnamefont {R.~D.}\ \bibnamefont {Johnson}}, \bibinfo {author} {\bibfnamefont {I.}~\bibnamefont {Kimchi}}, \bibinfo {author} {\bibfnamefont {R.}~\bibnamefont {Morris}}, \bibinfo {author} {\bibfnamefont {A.}~\bibnamefont {Bombardi}}, \bibinfo {author} {\bibfnamefont {J.~G.}\ \bibnamefont {Analytis}}, \bibinfo {author} {\bibfnamefont {A.}~\bibnamefont {Vishwanath}},\ and\ \bibinfo {author} {\bibfnamefont {R.}~\bibnamefont {Coldea}},\ }\bibfield  {title} {\bibinfo {title} {{Noncoplanar and counterrotating incommensurate magnetic order stabilized by Kitaev interactions in {$\gamma$}-Li$_2$IrO$_3$}},\ }\bibfield  {journal} {\bibinfo  {journal} {Physical Review Letters}\ }\textbf {\bibinfo {volume} {113}},\ \href {https://doi.org/10.1103/PhysRevLett.113.197201} {10.1103/PhysRevLett.113.197201} (\bibinfo {year} {2014})\BibitemShut {NoStop}%
\bibitem [{\citenamefont {Freund}\ \emph {et~al.}(2016)\citenamefont {Freund}, \citenamefont {Williams}, \citenamefont {Johnson}, \citenamefont {Coldea}, \citenamefont {Gegenwart},\ and\ \citenamefont {Jesche}}]{Freund2016SingleOxides}%
  \BibitemOpen
  \bibfield  {author} {\bibinfo {author} {\bibfnamefont {F.}~\bibnamefont {Freund}}, \bibinfo {author} {\bibfnamefont {S.~C.}\ \bibnamefont {Williams}}, \bibinfo {author} {\bibfnamefont {R.~D.}\ \bibnamefont {Johnson}}, \bibinfo {author} {\bibfnamefont {R.}~\bibnamefont {Coldea}}, \bibinfo {author} {\bibfnamefont {P.}~\bibnamefont {Gegenwart}},\ and\ \bibinfo {author} {\bibfnamefont {A.}~\bibnamefont {Jesche}},\ }\bibfield  {title} {\bibinfo {title} {{Single crystal growth from separated educts and its application to lithium transition-metal oxides}},\ }\bibfield  {journal} {\bibinfo  {journal} {Scientific Reports}\ }\textbf {\bibinfo {volume} {6}},\ \href {https://doi.org/10.1038/srep35362} {10.1038/srep35362} (\bibinfo {year} {2016})\BibitemShut {NoStop}%
\bibitem [{\citenamefont {Halloran}\ \emph {et~al.}(2022)\citenamefont {Halloran}, \citenamefont {Wang}, \citenamefont {Li}, \citenamefont {Rousochatzakis}, \citenamefont {Chauhan}, \citenamefont {Stone}, \citenamefont {Takayama}, \citenamefont {Takagi}, \citenamefont {Armitage}, \citenamefont {Perkins},\ and\ \citenamefont {Broholm}}]{HalloranLi2IrO3}%
  \BibitemOpen
  \bibfield  {author} {\bibinfo {author} {\bibfnamefont {T.}~\bibnamefont {Halloran}}, \bibinfo {author} {\bibfnamefont {Y.}~\bibnamefont {Wang}}, \bibinfo {author} {\bibfnamefont {M.}~\bibnamefont {Li}}, \bibinfo {author} {\bibfnamefont {I.}~\bibnamefont {Rousochatzakis}}, \bibinfo {author} {\bibfnamefont {P.}~\bibnamefont {Chauhan}}, \bibinfo {author} {\bibfnamefont {M.~B.}\ \bibnamefont {Stone}}, \bibinfo {author} {\bibfnamefont {T.}~\bibnamefont {Takayama}}, \bibinfo {author} {\bibfnamefont {H.}~\bibnamefont {Takagi}}, \bibinfo {author} {\bibfnamefont {N.~P.}\ \bibnamefont {Armitage}}, \bibinfo {author} {\bibfnamefont {N.~B.}\ \bibnamefont {Perkins}},\ and\ \bibinfo {author} {\bibfnamefont {C.}~\bibnamefont {Broholm}},\ }\bibfield  {title} {\bibinfo {title} {Magnetic excitations and interactions in the kitaev hyperhoneycomb iridate $\ensuremath{\beta}\text{\ensuremath{-}}{\mathrm{li}}_{2}{\mathrm{iro}}_{3}$},\ }\href {https://doi.org/10.1103/PhysRevB.106.064423} {\bibfield  {journal} {\bibinfo  {journal}
  {Phys. Rev. B}\ }\textbf {\bibinfo {volume} {106}},\ \bibinfo {pages} {064423} (\bibinfo {year} {2022})}\BibitemShut {NoStop}%
\bibitem [{\citenamefont {Bette}\ \emph {et~al.}(2017)\citenamefont {Bette}, \citenamefont {Takayama}, \citenamefont {Kitagawa}, \citenamefont {Takano}, \citenamefont {Takagi},\ and\ \citenamefont {Dinnebier}}]{Bette2017SolutionH3LiIr2O6}%
  \BibitemOpen
  \bibfield  {author} {\bibinfo {author} {\bibfnamefont {S.}~\bibnamefont {Bette}}, \bibinfo {author} {\bibfnamefont {T.}~\bibnamefont {Takayama}}, \bibinfo {author} {\bibfnamefont {K.}~\bibnamefont {Kitagawa}}, \bibinfo {author} {\bibfnamefont {R.}~\bibnamefont {Takano}}, \bibinfo {author} {\bibfnamefont {H.}~\bibnamefont {Takagi}},\ and\ \bibinfo {author} {\bibfnamefont {R.~E.}\ \bibnamefont {Dinnebier}},\ }\bibfield  {title} {\bibinfo {title} {{Solution of the heavily stacking faulted crystal structure of the honeycomb iridate H$_3$LiIr$_2$O$_6$}},\ }\href {https://doi.org/10.1039/c7dt02978k} {\bibfield  {journal} {\bibinfo  {journal} {Dalton Transactions}\ }\textbf {\bibinfo {volume} {46}},\ \bibinfo {pages} {15216} (\bibinfo {year} {2017})}\BibitemShut {NoStop}%
\bibitem [{\citenamefont {Kitagawa}\ \emph {et~al.}(2018)\citenamefont {Kitagawa}, \citenamefont {Takayama}, \citenamefont {Matsumoto}, \citenamefont {Kato}, \citenamefont {Takano}, \citenamefont {Kishimoto}, \citenamefont {Bette}, \citenamefont {Dinnebier}, \citenamefont {Jackeli},\ and\ \citenamefont {Takagi}}]{Kitagawa2018ALattice}%
  \BibitemOpen
  \bibfield  {author} {\bibinfo {author} {\bibfnamefont {K.}~\bibnamefont {Kitagawa}}, \bibinfo {author} {\bibfnamefont {T.}~\bibnamefont {Takayama}}, \bibinfo {author} {\bibfnamefont {Y.}~\bibnamefont {Matsumoto}}, \bibinfo {author} {\bibfnamefont {A.}~\bibnamefont {Kato}}, \bibinfo {author} {\bibfnamefont {R.}~\bibnamefont {Takano}}, \bibinfo {author} {\bibfnamefont {Y.}~\bibnamefont {Kishimoto}}, \bibinfo {author} {\bibfnamefont {S.}~\bibnamefont {Bette}}, \bibinfo {author} {\bibfnamefont {R.}~\bibnamefont {Dinnebier}}, \bibinfo {author} {\bibfnamefont {G.}~\bibnamefont {Jackeli}},\ and\ \bibinfo {author} {\bibfnamefont {H.}~\bibnamefont {Takagi}},\ }\bibfield  {title} {\bibinfo {title} {{A spin-orbital-entangled quantum liquid on a honeycomb lattice}},\ }\href {https://doi.org/10.1038/nature25482} {\bibfield  {journal} {\bibinfo  {journal} {Nature}\ }\textbf {\bibinfo {volume} {554}},\ \bibinfo {pages} {341} (\bibinfo {year} {2018})}\BibitemShut {NoStop}%
\bibitem [{\citenamefont {Geirhos}\ \emph {et~al.}(2020)\citenamefont {Geirhos}, \citenamefont {Lunkenheimer}, \citenamefont {Blankenhorn}, \citenamefont {Claus}, \citenamefont {Matsumoto}, \citenamefont {Kitagawa}, \citenamefont {Takayama}, \citenamefont {Takagi}, \citenamefont {K{\'{e}}zsm{\'{a}}rki},\ and\ \citenamefont {Loidl}}]{Geirhos2020QuantumO6}%
  \BibitemOpen
  \bibfield  {author} {\bibinfo {author} {\bibfnamefont {K.}~\bibnamefont {Geirhos}}, \bibinfo {author} {\bibfnamefont {P.}~\bibnamefont {Lunkenheimer}}, \bibinfo {author} {\bibfnamefont {M.}~\bibnamefont {Blankenhorn}}, \bibinfo {author} {\bibfnamefont {R.}~\bibnamefont {Claus}}, \bibinfo {author} {\bibfnamefont {Y.}~\bibnamefont {Matsumoto}}, \bibinfo {author} {\bibfnamefont {K.}~\bibnamefont {Kitagawa}}, \bibinfo {author} {\bibfnamefont {T.}~\bibnamefont {Takayama}}, \bibinfo {author} {\bibfnamefont {H.}~\bibnamefont {Takagi}}, \bibinfo {author} {\bibfnamefont {I.}~\bibnamefont {K{\'{e}}zsm{\'{a}}rki}},\ and\ \bibinfo {author} {\bibfnamefont {A.}~\bibnamefont {Loidl}},\ }\bibfield  {title} {\bibinfo {title} {{Quantum paraelectricity in the Kitaev quantum spin liquid candidates H$_3$LiIr$_2$O$_6$ and D$_3$LiIr$_2$O$_6$}},\ }\href {https://doi.org/10.1103/PhysRevB.101.184410} {\bibfield  {journal} {\bibinfo  {journal} {Physical Review B}\ }\textbf {\bibinfo {volume} {101}},\ \bibinfo {pages} {184410}
  (\bibinfo {year} {2020})}\BibitemShut {NoStop}%
\bibitem [{\citenamefont {Wang}\ \emph {et~al.}(2018)\citenamefont {Wang}, \citenamefont {Zhang},\ and\ \citenamefont {Wang}}]{Wang2018PossibleQP}%
  \BibitemOpen
  \bibfield  {author} {\bibinfo {author} {\bibfnamefont {S.}~\bibnamefont {Wang}}, \bibinfo {author} {\bibfnamefont {L.}~\bibnamefont {Zhang}},\ and\ \bibinfo {author} {\bibfnamefont {F.}~\bibnamefont {Wang}},\ }\bibfield  {title} {\bibinfo {title} {Possible quantum paraelectric state in kitaev spin liquid candidate h$_3$liir$_2$o$_6$},\ }\href@noop {} {\bibfield  {journal} {\bibinfo  {journal} {Science China Physics, Mechanics \& Astronomy}\ }\textbf {\bibinfo {volume} {63}} (\bibinfo {year} {2018})}\BibitemShut {NoStop}%
\bibitem [{\citenamefont {Yadav}\ \emph {et~al.}(2018)\citenamefont {Yadav}, \citenamefont {Ray}, \citenamefont {Eldeeb}, \citenamefont {Nishimoto}, \citenamefont {Hozoi},\ and\ \citenamefont {van~den Brink}}]{Yadav2018HLIOdisorder}%
  \BibitemOpen
  \bibfield  {author} {\bibinfo {author} {\bibfnamefont {R.}~\bibnamefont {Yadav}}, \bibinfo {author} {\bibfnamefont {R.}~\bibnamefont {Ray}}, \bibinfo {author} {\bibfnamefont {M.~S.}\ \bibnamefont {Eldeeb}}, \bibinfo {author} {\bibfnamefont {S.}~\bibnamefont {Nishimoto}}, \bibinfo {author} {\bibfnamefont {L.}~\bibnamefont {Hozoi}},\ and\ \bibinfo {author} {\bibfnamefont {J.}~\bibnamefont {van~den Brink}},\ }\bibfield  {title} {\bibinfo {title} {{Strong Effect of Hydrogen Order on Magnetic Kitaev Interactions in H$_3$LiIr$_2$O$_6$}},\ }\href {https://doi.org/10.1103/PhysRevLett.121.197203} {\bibfield  {journal} {\bibinfo  {journal} {Phys. Rev. Lett.}\ }\textbf {\bibinfo {volume} {121}},\ \bibinfo {pages} {197203} (\bibinfo {year} {2018})}\BibitemShut {NoStop}%
\bibitem [{\citenamefont {Li}\ \emph {et~al.}(2018)\citenamefont {Li}, \citenamefont {Winter},\ and\ \citenamefont {Valent{\'{i}}}}]{Li2018RoleO6}%
  \BibitemOpen
  \bibfield  {author} {\bibinfo {author} {\bibfnamefont {Y.}~\bibnamefont {Li}}, \bibinfo {author} {\bibfnamefont {S.~M.}\ \bibnamefont {Winter}},\ and\ \bibinfo {author} {\bibfnamefont {R.}~\bibnamefont {Valent{\'{i}}}},\ }\bibfield  {title} {\bibinfo {title} {{Role of Hydrogen in the Spin-Orbital-Entangled Quantum Liquid Candidate H$_3$LiIr$_2$O$_6$}},\ }\bibfield  {journal} {\bibinfo  {journal} {Physical Review Letters}\ }\textbf {\bibinfo {volume} {121}},\ \href {https://doi.org/10.1103/PhysRevLett.121.247202} {10.1103/PhysRevLett.121.247202} (\bibinfo {year} {2018})\BibitemShut {NoStop}%
\bibitem [{\citenamefont {Knolle}\ \emph {et~al.}(2019)\citenamefont {Knolle}, \citenamefont {Moessner},\ and\ \citenamefont {Perkins}}]{Knolle2019Bond-DisorderedHopping}%
  \BibitemOpen
  \bibfield  {author} {\bibinfo {author} {\bibfnamefont {J.}~\bibnamefont {Knolle}}, \bibinfo {author} {\bibfnamefont {R.}~\bibnamefont {Moessner}},\ and\ \bibinfo {author} {\bibfnamefont {N.~B.}\ \bibnamefont {Perkins}},\ }\bibfield  {title} {\bibinfo {title} {{Bond-Disordered Spin Liquid and the Honeycomb Iridate H$_3$LiIr$_2$O$_6$ : Abundant Low-Energy Density of States from Random Majorana Hopping}},\ }\bibfield  {journal} {\bibinfo  {journal} {Physical Review Letters}\ }\textbf {\bibinfo {volume} {122}},\ \href {https://doi.org/10.1103/PhysRevLett.122.047202} {10.1103/PhysRevLett.122.047202} (\bibinfo {year} {2019})\BibitemShut {NoStop}%
\bibitem [{\citenamefont {Slagle}\ \emph {et~al.}(2018)\citenamefont {Slagle}, \citenamefont {Choi}, \citenamefont {Chern},\ and\ \citenamefont {Kim}}]{Slagle2018TheoryO6}%
  \BibitemOpen
  \bibfield  {author} {\bibinfo {author} {\bibfnamefont {K.}~\bibnamefont {Slagle}}, \bibinfo {author} {\bibfnamefont {W.}~\bibnamefont {Choi}}, \bibinfo {author} {\bibfnamefont {L.~E.}\ \bibnamefont {Chern}},\ and\ \bibinfo {author} {\bibfnamefont {Y.~B.}\ \bibnamefont {Kim}},\ }\bibfield  {title} {\bibinfo {title} {{Theory of a quantum spin liquid in the hydrogen-intercalated honeycomb iridate H$_3$LiIr$_2$O$_6$}},\ }\href {https://doi.org/10.1103/PhysRevB.97.115159} {\bibfield  {journal} {\bibinfo  {journal} {Physical Review B}\ }\textbf {\bibinfo {volume} {97}},\ \bibinfo {pages} {115159} (\bibinfo {year} {2018})}\BibitemShut {NoStop}%
\bibitem [{\citenamefont {Kimchi}\ \emph {et~al.}(2018)\citenamefont {Kimchi}, \citenamefont {Sheckelton}, \citenamefont {McQueen},\ and\ \citenamefont {Lee}}]{Kimchi2018ScalingSystems}%
  \BibitemOpen
  \bibfield  {author} {\bibinfo {author} {\bibfnamefont {I.}~\bibnamefont {Kimchi}}, \bibinfo {author} {\bibfnamefont {J.~P.}\ \bibnamefont {Sheckelton}}, \bibinfo {author} {\bibfnamefont {T.~M.}\ \bibnamefont {McQueen}},\ and\ \bibinfo {author} {\bibfnamefont {P.~A.}\ \bibnamefont {Lee}},\ }\bibfield  {title} {\bibinfo {title} {{Scaling and data collapse from local moments in frustrated disordered quantum spin systems}},\ }\bibfield  {journal} {\bibinfo  {journal} {Nature Communications}\ }\textbf {\bibinfo {volume} {9}},\ \href {https://doi.org/10.1038/s41467-018-06800-2} {10.1038/s41467-018-06800-2} (\bibinfo {year} {2018})\BibitemShut {NoStop}%
\bibitem [{\citenamefont {Lee}\ \emph {et~al.}(2023)\citenamefont {Lee}, \citenamefont {Lee}, \citenamefont {Choi}, \citenamefont {Wang}, \citenamefont {Luetkens}, \citenamefont {Shiroka}, \citenamefont {Jang}, \citenamefont {Yoon},\ and\ \citenamefont {Choi}}]{LeeChanhyeonRandomSingletHLIO2023}%
  \BibitemOpen
  \bibfield  {author} {\bibinfo {author} {\bibfnamefont {C.}~\bibnamefont {Lee}}, \bibinfo {author} {\bibfnamefont {S.}~\bibnamefont {Lee}}, \bibinfo {author} {\bibfnamefont {Y.}~\bibnamefont {Choi}}, \bibinfo {author} {\bibfnamefont {C.}~\bibnamefont {Wang}}, \bibinfo {author} {\bibfnamefont {H.}~\bibnamefont {Luetkens}}, \bibinfo {author} {\bibfnamefont {T.}~\bibnamefont {Shiroka}}, \bibinfo {author} {\bibfnamefont {Z.}~\bibnamefont {Jang}}, \bibinfo {author} {\bibfnamefont {Y.-G.}\ \bibnamefont {Yoon}},\ and\ \bibinfo {author} {\bibfnamefont {K.-Y.}\ \bibnamefont {Choi}},\ }\bibfield  {title} {\bibinfo {title} {{Coexistence of random singlets and disordered Kitaev spin liquid in H$_3$LiIr$_2$O$_6$}},\ }\href {https://doi.org/10.1103/PhysRevB.107.014424} {\bibfield  {journal} {\bibinfo  {journal} {Phys. Rev. B}\ }\textbf {\bibinfo {volume} {107}},\ \bibinfo {pages} {014424} (\bibinfo {year} {2023})}\BibitemShut {NoStop}%
\bibitem [{\citenamefont {Pei}\ \emph {et~al.}(2020)\citenamefont {Pei}, \citenamefont {Huang}, \citenamefont {Li}, \citenamefont {Chen}, \citenamefont {Xi}, \citenamefont {Wang}, \citenamefont {Shi}, \citenamefont {Yu}, \citenamefont {Liu}, \citenamefont {Wang}, \citenamefont {Ye}, \citenamefont {Huang},\ and\ \citenamefont {Mei}}]{Pei2020MagneticO6}%
  \BibitemOpen
  \bibfield  {author} {\bibinfo {author} {\bibfnamefont {S.}~\bibnamefont {Pei}}, \bibinfo {author} {\bibfnamefont {L.~L.}\ \bibnamefont {Huang}}, \bibinfo {author} {\bibfnamefont {G.}~\bibnamefont {Li}}, \bibinfo {author} {\bibfnamefont {X.}~\bibnamefont {Chen}}, \bibinfo {author} {\bibfnamefont {B.}~\bibnamefont {Xi}}, \bibinfo {author} {\bibfnamefont {X.~W.}\ \bibnamefont {Wang}}, \bibinfo {author} {\bibfnamefont {Y.}~\bibnamefont {Shi}}, \bibinfo {author} {\bibfnamefont {D.}~\bibnamefont {Yu}}, \bibinfo {author} {\bibfnamefont {C.}~\bibnamefont {Liu}}, \bibinfo {author} {\bibfnamefont {L.}~\bibnamefont {Wang}}, \bibinfo {author} {\bibfnamefont {F.}~\bibnamefont {Ye}}, \bibinfo {author} {\bibfnamefont {M.}~\bibnamefont {Huang}},\ and\ \bibinfo {author} {\bibfnamefont {J.~W.}\ \bibnamefont {Mei}},\ }\bibfield  {title} {\bibinfo {title} {{Magnetic Raman continuum in single-crystalline H$_3$LiIr$_2$O$_6$}},\ }\href {https://doi.org/10.1103/PhysRevB.101.201101} {\bibfield  {journal} {\bibinfo  {journal}
  {Physical Review B}\ }\textbf {\bibinfo {volume} {101}},\ \bibinfo {pages} {201101} (\bibinfo {year} {2020})}\BibitemShut {NoStop}%
\bibitem [{\citenamefont {De~la Torre}\ \emph {et~al.}(2023)\citenamefont {De~la Torre}, \citenamefont {Zager}, \citenamefont {Bahrami}, \citenamefont {Upton}, \citenamefont {Kim}, \citenamefont {Fabbris}, \citenamefont {Lee}, \citenamefont {Yang}, \citenamefont {Haskel}, \citenamefont {Tafti} \emph {et~al.}}]{delatorre2023momentumindependent}%
  \BibitemOpen
  \bibfield  {author} {\bibinfo {author} {\bibfnamefont {A.}~\bibnamefont {De~la Torre}}, \bibinfo {author} {\bibfnamefont {B.}~\bibnamefont {Zager}}, \bibinfo {author} {\bibfnamefont {F.}~\bibnamefont {Bahrami}}, \bibinfo {author} {\bibfnamefont {M.}~\bibnamefont {Upton}}, \bibinfo {author} {\bibfnamefont {J.}~\bibnamefont {Kim}}, \bibinfo {author} {\bibfnamefont {G.}~\bibnamefont {Fabbris}}, \bibinfo {author} {\bibfnamefont {G.-H.}\ \bibnamefont {Lee}}, \bibinfo {author} {\bibfnamefont {W.}~\bibnamefont {Yang}}, \bibinfo {author} {\bibfnamefont {D.}~\bibnamefont {Haskel}}, \bibinfo {author} {\bibfnamefont {F.}~\bibnamefont {Tafti}}, \emph {et~al.},\ }\bibfield  {title} {\bibinfo {title} {{Momentum-independent magnetic excitation continuum in the honeycomb iridate H$_3$LiIr$_2$O$_6$}},\ }\href@noop {} {\bibfield  {journal} {\bibinfo  {journal} {Nature Communications}\ }\textbf {\bibinfo {volume} {14}},\ \bibinfo {pages} {5018} (\bibinfo {year} {2023})}\BibitemShut {NoStop}%
\bibitem [{\citenamefont {Yang}\ \emph {et~al.}(2022)\citenamefont {Yang}, \citenamefont {Huang}, \citenamefont {Zhu}, \citenamefont {Chen}, \citenamefont {Wu}, \citenamefont {Ding}, \citenamefont {Tan}, \citenamefont {Biswas}, \citenamefont {Hillier}, \citenamefont {Shi} \emph {et~al.}}]{yang2022muon}%
  \BibitemOpen
  \bibfield  {author} {\bibinfo {author} {\bibfnamefont {Y.-X.}\ \bibnamefont {Yang}}, \bibinfo {author} {\bibfnamefont {L.-L.}\ \bibnamefont {Huang}}, \bibinfo {author} {\bibfnamefont {Z.-H.}\ \bibnamefont {Zhu}}, \bibinfo {author} {\bibfnamefont {C.-S.}\ \bibnamefont {Chen}}, \bibinfo {author} {\bibfnamefont {Q.}~\bibnamefont {Wu}}, \bibinfo {author} {\bibfnamefont {Z.-F.}\ \bibnamefont {Ding}}, \bibinfo {author} {\bibfnamefont {C.}~\bibnamefont {Tan}}, \bibinfo {author} {\bibfnamefont {P.~K.}\ \bibnamefont {Biswas}}, \bibinfo {author} {\bibfnamefont {A.~D.}\ \bibnamefont {Hillier}}, \bibinfo {author} {\bibfnamefont {Y.-G.}\ \bibnamefont {Shi}}, \emph {et~al.},\ }\bibfield  {title} {\bibinfo {title} {{Muon Spin Relaxation Study of Spin Dynamics in Quantum Spin Liquid Candidate H$_3$LiIr$_2$O$_6$}},\ }\href@noop {} {\bibfield  {journal} {\bibinfo  {journal} {arXiv preprint arXiv:2201.12978}\ } (\bibinfo {year} {2022})}\BibitemShut {NoStop}%
\bibitem [{\citenamefont {Knolle}\ \emph {et~al.}(2014{\natexlab{a}})\citenamefont {Knolle}, \citenamefont {Chern}, \citenamefont {Kovrizhin}, \citenamefont {Moessner},\ and\ \citenamefont {Perkins}}]{knolle2014raman}%
  \BibitemOpen
  \bibfield  {author} {\bibinfo {author} {\bibfnamefont {J.}~\bibnamefont {Knolle}}, \bibinfo {author} {\bibfnamefont {G.-W.}\ \bibnamefont {Chern}}, \bibinfo {author} {\bibfnamefont {D.}~\bibnamefont {Kovrizhin}}, \bibinfo {author} {\bibfnamefont {R.}~\bibnamefont {Moessner}},\ and\ \bibinfo {author} {\bibfnamefont {N.}~\bibnamefont {Perkins}},\ }\bibfield  {title} {\bibinfo {title} {Raman scattering signatures of kitaev spin liquids in a$_2$iro$_3$ iridates with a=na or li},\ }\href@noop {} {\bibfield  {journal} {\bibinfo  {journal} {Physical review letters}\ }\textbf {\bibinfo {volume} {113}},\ \bibinfo {pages} {187201} (\bibinfo {year} {2014}{\natexlab{a}})}\BibitemShut {NoStop}%
\bibitem [{\citenamefont {Knolle}\ \emph {et~al.}(2014{\natexlab{b}})\citenamefont {Knolle}, \citenamefont {Kovrizhin}, \citenamefont {Chalker},\ and\ \citenamefont {Moessner}}]{Knolle2014DynamicsFluxes}%
  \BibitemOpen
  \bibfield  {author} {\bibinfo {author} {\bibfnamefont {J.}~\bibnamefont {Knolle}}, \bibinfo {author} {\bibfnamefont {D.~L.}\ \bibnamefont {Kovrizhin}}, \bibinfo {author} {\bibfnamefont {J.~T.}\ \bibnamefont {Chalker}},\ and\ \bibinfo {author} {\bibfnamefont {R.}~\bibnamefont {Moessner}},\ }\bibfield  {title} {\bibinfo {title} {{Dynamics of a two-dimensional quantum spin liquid: Signatures of emergent majorana fermions and fluxes}},\ }\href {https://doi.org/10.1103/PhysRevLett.112.207203} {\bibfield  {journal} {\bibinfo  {journal} {Physical Review Letters}\ }\textbf {\bibinfo {volume} {112}},\ \bibinfo {pages} {207203} (\bibinfo {year} {2014}{\natexlab{b}})}\BibitemShut {NoStop}%
\bibitem [{\citenamefont {Nasu}\ and\ \citenamefont {Motome}(2021)}]{Nasu2021}%
  \BibitemOpen
  \bibfield  {author} {\bibinfo {author} {\bibfnamefont {J.}~\bibnamefont {Nasu}}\ and\ \bibinfo {author} {\bibfnamefont {Y.}~\bibnamefont {Motome}},\ }\bibfield  {title} {\bibinfo {title} {Spin dynamics in the kitaev model with disorder: Quantum monte carlo study of dynamical spin structure factor, magnetic susceptibility, and nmr relaxation rate},\ }\href {https://doi.org/10.1103/PhysRevB.104.035116} {\bibfield  {journal} {\bibinfo  {journal} {Phys. Rev. B}\ }\textbf {\bibinfo {volume} {104}},\ \bibinfo {pages} {035116} (\bibinfo {year} {2021})}\BibitemShut {NoStop}%
\bibitem [{\citenamefont {Samarakoon}\ \emph {et~al.}(2017)\citenamefont {Samarakoon}, \citenamefont {Banerjee}, \citenamefont {Zhang}, \citenamefont {Kamiya}, \citenamefont {Nagler}, \citenamefont {Tennant}, \citenamefont {Lee},\ and\ \citenamefont {Batista}}]{Samarakoon2017}%
  \BibitemOpen
  \bibfield  {author} {\bibinfo {author} {\bibfnamefont {A.~M.}\ \bibnamefont {Samarakoon}}, \bibinfo {author} {\bibfnamefont {A.}~\bibnamefont {Banerjee}}, \bibinfo {author} {\bibfnamefont {S.-S.}\ \bibnamefont {Zhang}}, \bibinfo {author} {\bibfnamefont {Y.}~\bibnamefont {Kamiya}}, \bibinfo {author} {\bibfnamefont {S.~E.}\ \bibnamefont {Nagler}}, \bibinfo {author} {\bibfnamefont {D.~A.}\ \bibnamefont {Tennant}}, \bibinfo {author} {\bibfnamefont {S.-H.}\ \bibnamefont {Lee}},\ and\ \bibinfo {author} {\bibfnamefont {C.~D.}\ \bibnamefont {Batista}},\ }\bibfield  {title} {\bibinfo {title} {Comprehensive study of the dynamics of a classical kitaev spin liquid},\ }\href {https://doi.org/10.1103/PhysRevB.96.134408} {\bibfield  {journal} {\bibinfo  {journal} {Phys. Rev. B}\ }\textbf {\bibinfo {volume} {96}},\ \bibinfo {pages} {134408} (\bibinfo {year} {2017})}\BibitemShut {NoStop}%
\bibitem [{\citenamefont {Stone}\ \emph {et~al.}(2014)\citenamefont {Stone}, \citenamefont {Niedziela}, \citenamefont {Abernathy}, \citenamefont {Debeer-Schmitt}, \citenamefont {Ehlers}, \citenamefont {Garlea}, \citenamefont {Granroth}, \citenamefont {Graves-Brook}, \citenamefont {Kolesnikov}, \citenamefont {Podlesnyak},\ and\ \citenamefont {Winn}}]{Stone2014ASource}%
  \BibitemOpen
  \bibfield  {author} {\bibinfo {author} {\bibfnamefont {M.~B.}\ \bibnamefont {Stone}}, \bibinfo {author} {\bibfnamefont {J.~L.}\ \bibnamefont {Niedziela}}, \bibinfo {author} {\bibfnamefont {D.~L.}\ \bibnamefont {Abernathy}}, \bibinfo {author} {\bibfnamefont {L.}~\bibnamefont {Debeer-Schmitt}}, \bibinfo {author} {\bibfnamefont {G.}~\bibnamefont {Ehlers}}, \bibinfo {author} {\bibfnamefont {O.}~\bibnamefont {Garlea}}, \bibinfo {author} {\bibfnamefont {G.~E.}\ \bibnamefont {Granroth}}, \bibinfo {author} {\bibfnamefont {M.}~\bibnamefont {Graves-Brook}}, \bibinfo {author} {\bibfnamefont {A.~I.}\ \bibnamefont {Kolesnikov}}, \bibinfo {author} {\bibfnamefont {A.}~\bibnamefont {Podlesnyak}},\ and\ \bibinfo {author} {\bibfnamefont {B.}~\bibnamefont {Winn}},\ }\bibfield  {title} {\bibinfo {title} {{A comparison of four direct geometry time-of-flight spectrometers at the Spallation Neutron Source}},\ }\href {https://doi.org/10.1063/1.4870050} {\bibfield  {journal} {\bibinfo  {journal} {Review of Scientific Instruments}\
  }\textbf {\bibinfo {volume} {85}},\ \bibinfo {pages} {045113} (\bibinfo {year} {2014})}\BibitemShut {NoStop}%
\bibitem [{\citenamefont {Rodriguez}\ \emph {et~al.}(2008)\citenamefont {Rodriguez}, \citenamefont {Adler}, \citenamefont {Brand}, \citenamefont {Broholm}, \citenamefont {Cook}, \citenamefont {Brocker}, \citenamefont {Hammond}, \citenamefont {Huang}, \citenamefont {Hundertmark}, \citenamefont {Lynn}, \citenamefont {Maliszewskyj}, \citenamefont {Moyer}, \citenamefont {Orndorff}, \citenamefont {Pierce}, \citenamefont {Pike}, \citenamefont {Scharfstein}, \citenamefont {Smee},\ and\ \citenamefont {Vilaseca}}]{Rodriguez2008MACSNIST}%
  \BibitemOpen
  \bibfield  {author} {\bibinfo {author} {\bibfnamefont {J.~A.}\ \bibnamefont {Rodriguez}}, \bibinfo {author} {\bibfnamefont {D.~M.}\ \bibnamefont {Adler}}, \bibinfo {author} {\bibfnamefont {P.~C.}\ \bibnamefont {Brand}}, \bibinfo {author} {\bibfnamefont {C.}~\bibnamefont {Broholm}}, \bibinfo {author} {\bibfnamefont {J.~C.}\ \bibnamefont {Cook}}, \bibinfo {author} {\bibfnamefont {C.}~\bibnamefont {Brocker}}, \bibinfo {author} {\bibfnamefont {R.}~\bibnamefont {Hammond}}, \bibinfo {author} {\bibfnamefont {Z.}~\bibnamefont {Huang}}, \bibinfo {author} {\bibfnamefont {P.}~\bibnamefont {Hundertmark}}, \bibinfo {author} {\bibfnamefont {J.~W.}\ \bibnamefont {Lynn}}, \bibinfo {author} {\bibfnamefont {N.~C.}\ \bibnamefont {Maliszewskyj}}, \bibinfo {author} {\bibfnamefont {J.}~\bibnamefont {Moyer}}, \bibinfo {author} {\bibfnamefont {J.}~\bibnamefont {Orndorff}}, \bibinfo {author} {\bibfnamefont {D.}~\bibnamefont {Pierce}}, \bibinfo {author} {\bibfnamefont {T.~D.}\ \bibnamefont {Pike}}, \bibinfo {author} {\bibfnamefont
  {G.}~\bibnamefont {Scharfstein}}, \bibinfo {author} {\bibfnamefont {S.~A.}\ \bibnamefont {Smee}},\ and\ \bibinfo {author} {\bibfnamefont {R.}~\bibnamefont {Vilaseca}},\ }\bibfield  {title} {\bibinfo {title} {{MACS - A new high intensity cold neutron spectrometer at NIST}},\ }in\ \href {https://doi.org/10.1088/0957-0233/19/3/034023} {\emph {\bibinfo {booktitle} {Measurement Science and Technology}}},\ Vol.~\bibinfo {volume} {19}\ (\bibinfo  {publisher} {Institute of Physics Publishing},\ \bibinfo {year} {2008})\ p.\ \bibinfo {pages} {034023}\BibitemShut {NoStop}%
\bibitem [{\citenamefont {Laurita}(2017)}]{Laurita2017LowMagnets}%
  \BibitemOpen
  \bibfield  {author} {\bibinfo {author} {\bibfnamefont {N.~J.}\ \bibnamefont {Laurita}},\ }\emph {\bibinfo {title} {{Low Energy Electrodynamics of Quantum Magnets}}},\ \href@noop {} {Ph.D. thesis},\ \bibinfo  {school} {Johns Hopkins University} (\bibinfo {year} {2017})\BibitemShut {NoStop}%
\bibitem [{\citenamefont {Sch{\"a}rpf}\ and\ \citenamefont {Capellmann}(1993)}]{Scharpf1993TheXM}%
  \BibitemOpen
  \bibfield  {author} {\bibinfo {author} {\bibfnamefont {O.}~\bibnamefont {Sch{\"a}rpf}}\ and\ \bibinfo {author} {\bibfnamefont {H.}~\bibnamefont {Capellmann}},\ }\bibfield  {title} {\bibinfo {title} {The xyz-difference method with polarized neutrons and the separation of coherent, spin incoherent, and magnetic scattering cross sections in a multidetector†},\ }\href@noop {} {\bibfield  {journal} {\bibinfo  {journal} {Physica Status Solidi (a)}\ }\textbf {\bibinfo {volume} {135}},\ \bibinfo {pages} {359} (\bibinfo {year} {1993})}\BibitemShut {NoStop}%
\bibitem [{\citenamefont {Lovesey}(1984)}]{lovesey1984theory}%
  \BibitemOpen
  \bibfield  {author} {\bibinfo {author} {\bibfnamefont {S.~W.}\ \bibnamefont {Lovesey}},\ }\href@noop {} {\emph {\bibinfo {title} {Theory of neutron scattering from condensed matter}}},\ Vol.~\bibinfo {volume} {72}\ (\bibinfo  {publisher} {Clarendon Press},\ \bibinfo {year} {1984})\BibitemShut {NoStop}%
\bibitem [{\citenamefont {Singh}\ \emph {et~al.}(2012)\citenamefont {Singh}, \citenamefont {Manni}, \citenamefont {Reuther}, \citenamefont {Berlijn}, \citenamefont {Thomale}, \citenamefont {Ku}, \citenamefont {Trebst},\ and\ \citenamefont {Gegenwart}}]{Singh2012Relevance3}%
  \BibitemOpen
  \bibfield  {author} {\bibinfo {author} {\bibfnamefont {Y.}~\bibnamefont {Singh}}, \bibinfo {author} {\bibfnamefont {S.}~\bibnamefont {Manni}}, \bibinfo {author} {\bibfnamefont {J.}~\bibnamefont {Reuther}}, \bibinfo {author} {\bibfnamefont {T.}~\bibnamefont {Berlijn}}, \bibinfo {author} {\bibfnamefont {R.}~\bibnamefont {Thomale}}, \bibinfo {author} {\bibfnamefont {W.}~\bibnamefont {Ku}}, \bibinfo {author} {\bibfnamefont {S.}~\bibnamefont {Trebst}},\ and\ \bibinfo {author} {\bibfnamefont {P.}~\bibnamefont {Gegenwart}},\ }\bibfield  {title} {\bibinfo {title} {{Relevance of the Heisenberg-Kitaev model for the honeycomb lattice iridates A$_2$IrO$_3$}},\ }\bibfield  {journal} {\bibinfo  {journal} {Physical Review Letters}\ }\textbf {\bibinfo {volume} {108}},\ \href {https://doi.org/10.1103/PhysRevLett.108.127203} {10.1103/PhysRevLett.108.127203} (\bibinfo {year} {2012})\BibitemShut {NoStop}%
\bibitem [{\citenamefont {Choi}\ \emph {et~al.}(2012)\citenamefont {Choi}, \citenamefont {Coldea}, \citenamefont {Kolmogorov}, \citenamefont {Lancaster}, \citenamefont {Mazin}, \citenamefont {Blundell}, \citenamefont {Radaelli}, \citenamefont {Singh}, \citenamefont {Gegenwart}, \citenamefont {Choi}, \citenamefont {Cheong}, \citenamefont {Baker}, \citenamefont {Stock},\ and\ \citenamefont {Taylor}}]{Choi2012Spin3}%
  \BibitemOpen
  \bibfield  {author} {\bibinfo {author} {\bibfnamefont {S.~K.}\ \bibnamefont {Choi}}, \bibinfo {author} {\bibfnamefont {R.}~\bibnamefont {Coldea}}, \bibinfo {author} {\bibfnamefont {A.~N.}\ \bibnamefont {Kolmogorov}}, \bibinfo {author} {\bibfnamefont {T.}~\bibnamefont {Lancaster}}, \bibinfo {author} {\bibfnamefont {I.~I.}\ \bibnamefont {Mazin}}, \bibinfo {author} {\bibfnamefont {S.~J.}\ \bibnamefont {Blundell}}, \bibinfo {author} {\bibfnamefont {P.~G.}\ \bibnamefont {Radaelli}}, \bibinfo {author} {\bibfnamefont {Y.}~\bibnamefont {Singh}}, \bibinfo {author} {\bibfnamefont {P.}~\bibnamefont {Gegenwart}}, \bibinfo {author} {\bibfnamefont {K.~R.}\ \bibnamefont {Choi}}, \bibinfo {author} {\bibfnamefont {S.~W.}\ \bibnamefont {Cheong}}, \bibinfo {author} {\bibfnamefont {P.~J.}\ \bibnamefont {Baker}}, \bibinfo {author} {\bibfnamefont {C.}~\bibnamefont {Stock}},\ and\ \bibinfo {author} {\bibfnamefont {J.}~\bibnamefont {Taylor}},\ }\bibfield  {title} {\bibinfo {title} {{Spin waves and revised crystal structure of
  honeycomb iridate Na$_2$IrO$_3$}},\ }\bibfield  {journal} {\bibinfo  {journal} {Physical Review Letters}\ }\textbf {\bibinfo {volume} {108}},\ \href {https://doi.org/10.1103/PhysRevLett.108.127204} {10.1103/PhysRevLett.108.127204} (\bibinfo {year} {2012})\BibitemShut {NoStop}%
\bibitem [{\citenamefont {Winter}\ \emph {et~al.}(2016)\citenamefont {Winter}, \citenamefont {Li}, \citenamefont {Jeschke},\ and\ \citenamefont {Valent{\'{i}}}}]{Winter2016ChallengesScales}%
  \BibitemOpen
  \bibfield  {author} {\bibinfo {author} {\bibfnamefont {S.~M.}\ \bibnamefont {Winter}}, \bibinfo {author} {\bibfnamefont {Y.}~\bibnamefont {Li}}, \bibinfo {author} {\bibfnamefont {H.~O.}\ \bibnamefont {Jeschke}},\ and\ \bibinfo {author} {\bibfnamefont {R.}~\bibnamefont {Valent{\'{i}}}},\ }\bibfield  {title} {\bibinfo {title} {{Challenges in design of Kitaev materials: Magnetic interactions from competing energy scales}},\ }\href {https://doi.org/10.1103/PhysRevB.93.214431} {\bibfield  {journal} {\bibinfo  {journal} {Physical Review B}\ }\textbf {\bibinfo {volume} {93}},\ \bibinfo {pages} {214431} (\bibinfo {year} {2016})}\BibitemShut {NoStop}%
\bibitem [{\citenamefont {Kao}\ \emph {et~al.}(2021)\citenamefont {Kao}, \citenamefont {Knolle}, \citenamefont {Hal{\'a}sz}, \citenamefont {Moessner},\ and\ \citenamefont {Perkins}}]{Kao2020Vacancy-inducedLiquid}%
  \BibitemOpen
  \bibfield  {author} {\bibinfo {author} {\bibfnamefont {W.-H.}\ \bibnamefont {Kao}}, \bibinfo {author} {\bibfnamefont {J.}~\bibnamefont {Knolle}}, \bibinfo {author} {\bibfnamefont {G.~B.}\ \bibnamefont {Hal{\'a}sz}}, \bibinfo {author} {\bibfnamefont {R.}~\bibnamefont {Moessner}},\ and\ \bibinfo {author} {\bibfnamefont {N.~B.}\ \bibnamefont {Perkins}},\ }\bibfield  {title} {\bibinfo {title} {Vacancy-induced low-energy density of states in the kitaev spin liquid},\ }\href@noop {} {\bibfield  {journal} {\bibinfo  {journal} {Physical Review X}\ }\textbf {\bibinfo {volume} {11}},\ \bibinfo {pages} {011034} (\bibinfo {year} {2021})}\BibitemShut {NoStop}%
\bibitem [{\citenamefont {Kao}\ \emph {et~al.}(2023)\citenamefont {Kao}, \citenamefont {Perkins},\ and\ \citenamefont {Hal{\'a}sz}}]{kao2023vacancy}%
  \BibitemOpen
  \bibfield  {author} {\bibinfo {author} {\bibfnamefont {W.-H.}\ \bibnamefont {Kao}}, \bibinfo {author} {\bibfnamefont {N.~B.}\ \bibnamefont {Perkins}},\ and\ \bibinfo {author} {\bibfnamefont {G.~B.}\ \bibnamefont {Hal{\'a}sz}},\ }\bibfield  {title} {\bibinfo {title} {Vacancy spectroscopy of non-abelian kitaev spin liquids},\ }\href@noop {} {\bibfield  {journal} {\bibinfo  {journal} {arXiv preprint arXiv:2307.10376}\ } (\bibinfo {year} {2023})}\BibitemShut {NoStop}%
\end{thebibliography}%


\begin{thebibliography}{12}%
\makeatletter
\providecommand \@ifxundefined [1]{%
 \@ifx{#1\undefined}
}%
\providecommand \@ifnum [1]{%
 \ifnum #1\expandafter \@firstoftwo
 \else \expandafter \@secondoftwo
 \fi
}%
\providecommand \@ifx [1]{%
 \ifx #1\expandafter \@firstoftwo
 \else \expandafter \@secondoftwo
 \fi
}%
\providecommand \natexlab [1]{#1}%
\providecommand \enquote  [1]{``#1''}%
\providecommand \bibnamefont  [1]{#1}%
\providecommand \bibfnamefont [1]{#1}%
\providecommand \citenamefont [1]{#1}%
\providecommand \href@noop [0]{\@secondoftwo}%
\providecommand \href [0]{\begingroup \@sanitize@url \@href}%
\providecommand \@href[1]{\@@startlink{#1}\@@href}%
\providecommand \@@href[1]{\endgroup#1\@@endlink}%
\providecommand \@sanitize@url [0]{\catcode `\\12\catcode `\$12\catcode
  `\&12\catcode `\#12\catcode `\^12\catcode `\_12\catcode `\%12\relax}%
\providecommand \@@startlink[1]{}%
\providecommand \@@endlink[0]{}%
\providecommand \url  [0]{\begingroup\@sanitize@url \@url }%
\providecommand \@url [1]{\endgroup\@href {#1}{\urlprefix }}%
\providecommand \urlprefix  [0]{URL }%
\providecommand \Eprint [0]{\href }%
\providecommand \doibase [0]{https://doi.org/}%
\providecommand \selectlanguage [0]{\@gobble}%
\providecommand \bibinfo  [0]{\@secondoftwo}%
\providecommand \bibfield  [0]{\@secondoftwo}%
\providecommand \translation [1]{[#1]}%
\providecommand \BibitemOpen [0]{}%
\providecommand \bibitemStop [0]{}%
\providecommand \bibitemNoStop [0]{.\EOS\space}%
\providecommand \EOS [0]{\spacefactor3000\relax}%
\providecommand \BibitemShut  [1]{\csname bibitem#1\endcsname}%
\let\auto@bib@innerbib\@empty
\bibitem [{\citenamefont {Rodriguez}\ \emph {et~al.}(2008)\citenamefont
  {Rodriguez}, \citenamefont {Adler}, \citenamefont {Brand}, \citenamefont
  {Broholm}, \citenamefont {Cook}, \citenamefont {Brocker}, \citenamefont
  {Hammond}, \citenamefont {Huang}, \citenamefont {Hundertmark}, \citenamefont
  {Lynn}, \citenamefont {Maliszewskyj}, \citenamefont {Moyer}, \citenamefont
  {Orndorff}, \citenamefont {Pierce}, \citenamefont {Pike}, \citenamefont
  {Scharfstein}, \citenamefont {Smee},\ and\ \citenamefont
  {Vilaseca}}]{Rodriguez2008MACSNIST}%
  \BibitemOpen
  \bibfield  {author} {\bibinfo {author} {\bibfnamefont {J.~A.}\ \bibnamefont
  {Rodriguez}}, \bibinfo {author} {\bibfnamefont {D.~M.}\ \bibnamefont
  {Adler}}, \bibinfo {author} {\bibfnamefont {P.~C.}\ \bibnamefont {Brand}},
  \bibinfo {author} {\bibfnamefont {C.}~\bibnamefont {Broholm}}, \bibinfo
  {author} {\bibfnamefont {J.~C.}\ \bibnamefont {Cook}}, \bibinfo {author}
  {\bibfnamefont {C.}~\bibnamefont {Brocker}}, \bibinfo {author} {\bibfnamefont
  {R.}~\bibnamefont {Hammond}}, \bibinfo {author} {\bibfnamefont
  {Z.}~\bibnamefont {Huang}}, \bibinfo {author} {\bibfnamefont
  {P.}~\bibnamefont {Hundertmark}}, \bibinfo {author} {\bibfnamefont {J.~W.}\
  \bibnamefont {Lynn}}, \bibinfo {author} {\bibfnamefont {N.~C.}\ \bibnamefont
  {Maliszewskyj}}, \bibinfo {author} {\bibfnamefont {J.}~\bibnamefont {Moyer}},
  \bibinfo {author} {\bibfnamefont {J.}~\bibnamefont {Orndorff}}, \bibinfo
  {author} {\bibfnamefont {D.}~\bibnamefont {Pierce}}, \bibinfo {author}
  {\bibfnamefont {T.~D.}\ \bibnamefont {Pike}}, \bibinfo {author}
  {\bibfnamefont {G.}~\bibnamefont {Scharfstein}}, \bibinfo {author}
  {\bibfnamefont {S.~A.}\ \bibnamefont {Smee}},\ and\ \bibinfo {author}
  {\bibfnamefont {R.}~\bibnamefont {Vilaseca}},\ }\bibfield  {title} {\bibinfo
  {title} {{MACS - A new high intensity cold neutron spectrometer at NIST}},\
  }in\ \href {https://doi.org/10.1088/0957-0233/19/3/034023} {\emph {\bibinfo
  {booktitle} {Measurement Science and Technology}}},\ Vol.~\bibinfo {volume}
  {19}\ (\bibinfo  {publisher} {Institute of Physics Publishing},\ \bibinfo
  {year} {2008})\ p.\ \bibinfo {pages} {034023}\BibitemShut {NoStop}%
\bibitem [{\citenamefont {Stone}\ \emph {et~al.}(2014)\citenamefont {Stone},
  \citenamefont {Niedziela}, \citenamefont {Abernathy}, \citenamefont
  {Debeer-Schmitt}, \citenamefont {Ehlers}, \citenamefont {Garlea},
  \citenamefont {Granroth}, \citenamefont {Graves-Brook}, \citenamefont
  {Kolesnikov}, \citenamefont {Podlesnyak},\ and\ \citenamefont
  {Winn}}]{Stone2014ASource}%
  \BibitemOpen
  \bibfield  {author} {\bibinfo {author} {\bibfnamefont {M.~B.}\ \bibnamefont
  {Stone}}, \bibinfo {author} {\bibfnamefont {J.~L.}\ \bibnamefont
  {Niedziela}}, \bibinfo {author} {\bibfnamefont {D.~L.}\ \bibnamefont
  {Abernathy}}, \bibinfo {author} {\bibfnamefont {L.}~\bibnamefont
  {Debeer-Schmitt}}, \bibinfo {author} {\bibfnamefont {G.}~\bibnamefont
  {Ehlers}}, \bibinfo {author} {\bibfnamefont {O.}~\bibnamefont {Garlea}},
  \bibinfo {author} {\bibfnamefont {G.~E.}\ \bibnamefont {Granroth}}, \bibinfo
  {author} {\bibfnamefont {M.}~\bibnamefont {Graves-Brook}}, \bibinfo {author}
  {\bibfnamefont {A.~I.}\ \bibnamefont {Kolesnikov}}, \bibinfo {author}
  {\bibfnamefont {A.}~\bibnamefont {Podlesnyak}},\ and\ \bibinfo {author}
  {\bibfnamefont {B.}~\bibnamefont {Winn}},\ }\bibfield  {title} {\bibinfo
  {title} {{A comparison of four direct geometry time-of-flight spectrometers
  at the Spallation Neutron Source}},\ }\href
  {https://doi.org/10.1063/1.4870050} {\bibfield  {journal} {\bibinfo
  {journal} {Review of Scientific Instruments}\ }\textbf {\bibinfo {volume}
  {85}},\ \bibinfo {pages} {045113} (\bibinfo {year} {2014})}\BibitemShut
  {NoStop}%
\bibitem [{\citenamefont {Kitagawa}\ \emph {et~al.}(2018)\citenamefont
  {Kitagawa}, \citenamefont {Takayama}, \citenamefont {Matsumoto},
  \citenamefont {Kato}, \citenamefont {Takano}, \citenamefont {Kishimoto},
  \citenamefont {Bette}, \citenamefont {Dinnebier}, \citenamefont {Jackeli},\
  and\ \citenamefont {Takagi}}]{Kitagawa2018ALattice}%
  \BibitemOpen
  \bibfield  {author} {\bibinfo {author} {\bibfnamefont {K.}~\bibnamefont
  {Kitagawa}}, \bibinfo {author} {\bibfnamefont {T.}~\bibnamefont {Takayama}},
  \bibinfo {author} {\bibfnamefont {Y.}~\bibnamefont {Matsumoto}}, \bibinfo
  {author} {\bibfnamefont {A.}~\bibnamefont {Kato}}, \bibinfo {author}
  {\bibfnamefont {R.}~\bibnamefont {Takano}}, \bibinfo {author} {\bibfnamefont
  {Y.}~\bibnamefont {Kishimoto}}, \bibinfo {author} {\bibfnamefont
  {S.}~\bibnamefont {Bette}}, \bibinfo {author} {\bibfnamefont
  {R.}~\bibnamefont {Dinnebier}}, \bibinfo {author} {\bibfnamefont
  {G.}~\bibnamefont {Jackeli}},\ and\ \bibinfo {author} {\bibfnamefont
  {H.}~\bibnamefont {Takagi}},\ }\bibfield  {title} {\bibinfo {title} {{A
  spin-orbital-entangled quantum liquid on a honeycomb lattice}},\ }\href
  {https://doi.org/10.1038/nature25482} {\bibfield  {journal} {\bibinfo
  {journal} {Nature}\ }\textbf {\bibinfo {volume} {554}},\ \bibinfo {pages}
  {341} (\bibinfo {year} {2018})}\BibitemShut {NoStop}%
\bibitem [{\citenamefont {Schmitt}\ and\ \citenamefont
  {Ouladdiaf}(1998)}]{Schmitt1998AbsorptionDiffraction}%
  \BibitemOpen
  \bibfield  {author} {\bibinfo {author} {\bibfnamefont {D.}~\bibnamefont
  {Schmitt}}\ and\ \bibinfo {author} {\bibfnamefont {B.}~\bibnamefont
  {Ouladdiaf}},\ }\bibfield  {title} {\bibinfo {title} {{Absorption Correction
  for Annular Cylindrical Samples in Powder Neutron Diffraction}},\ }\href
  {https://doi.org/10.1107/S0021889898002672} {\bibfield  {journal} {\bibinfo
  {journal} {Journal of Applied Crystallography}\ }\textbf {\bibinfo {volume}
  {31}},\ \bibinfo {pages} {620} (\bibinfo {year} {1998})}\BibitemShut
  {NoStop}%
\bibitem [{\citenamefont {Hong}\ \emph {et~al.}(2006)\citenamefont {Hong},
  \citenamefont {Kenzelmann}, \citenamefont {Turnbull}, \citenamefont {Landee},
  \citenamefont {Lewis}, \citenamefont {Schmidt}, \citenamefont {Uhrig},
  \citenamefont {Qiu}, \citenamefont {Broholm},\ and\ \citenamefont
  {Reich}}]{Hong2006NeutronParamagnet}%
  \BibitemOpen
  \bibfield  {author} {\bibinfo {author} {\bibfnamefont {T.}~\bibnamefont
  {Hong}}, \bibinfo {author} {\bibfnamefont {M.}~\bibnamefont {Kenzelmann}},
  \bibinfo {author} {\bibfnamefont {M.~M.}\ \bibnamefont {Turnbull}}, \bibinfo
  {author} {\bibfnamefont {C.~P.}\ \bibnamefont {Landee}}, \bibinfo {author}
  {\bibfnamefont {B.~D.}\ \bibnamefont {Lewis}}, \bibinfo {author}
  {\bibfnamefont {K.~P.}\ \bibnamefont {Schmidt}}, \bibinfo {author}
  {\bibfnamefont {G.~S.}\ \bibnamefont {Uhrig}}, \bibinfo {author}
  {\bibfnamefont {Y.}~\bibnamefont {Qiu}}, \bibinfo {author} {\bibfnamefont
  {C.}~\bibnamefont {Broholm}},\ and\ \bibinfo {author} {\bibfnamefont
  {D.}~\bibnamefont {Reich}},\ }\bibfield  {title} {\bibinfo {title} {Neutron
  scattering from a coordination polymer quantum paramagnet},\ }\href
  {https://doi.org/10.1103/PhysRevB.74.094434} {\bibfield  {journal} {\bibinfo
  {journal} {Phys. Rev. B}\ }\textbf {\bibinfo {volume} {74}},\ \bibinfo
  {pages} {094434} (\bibinfo {year} {2006})}\BibitemShut {NoStop}%
\bibitem [{\citenamefont {Azuah}\ \emph {et~al.}(2009)\citenamefont {Azuah},
  \citenamefont {Kneller}, \citenamefont {Qiu}, \citenamefont
  {Tregenna-Piggott}, \citenamefont {Brown}, \citenamefont {Copley},\ and\
  \citenamefont {Dimeo}}]{azuah2009dave}%
  \BibitemOpen
  \bibfield  {author} {\bibinfo {author} {\bibfnamefont {R.~T.}\ \bibnamefont
  {Azuah}}, \bibinfo {author} {\bibfnamefont {L.~R.}\ \bibnamefont {Kneller}},
  \bibinfo {author} {\bibfnamefont {Y.}~\bibnamefont {Qiu}}, \bibinfo {author}
  {\bibfnamefont {P.~L.}\ \bibnamefont {Tregenna-Piggott}}, \bibinfo {author}
  {\bibfnamefont {C.~M.}\ \bibnamefont {Brown}}, \bibinfo {author}
  {\bibfnamefont {J.~R.}\ \bibnamefont {Copley}},\ and\ \bibinfo {author}
  {\bibfnamefont {R.~M.}\ \bibnamefont {Dimeo}},\ }\bibfield  {title} {\bibinfo
  {title} {Dave: a comprehensive software suite for the reduction,
  visualization, and analysis of low energy neutron spectroscopic data},\
  }\href@noop {} {\bibfield  {journal} {\bibinfo  {journal} {Journal of
  research of the National Institute of Standards and Technology}\ }\textbf
  {\bibinfo {volume} {114}},\ \bibinfo {pages} {341} (\bibinfo {year}
  {2009})}\BibitemShut {NoStop}%
\bibitem [{\citenamefont {Nicklow}\ \emph {et~al.}(1972)\citenamefont
  {Nicklow}, \citenamefont {Wakabayashi},\ and\ \citenamefont
  {Smith}}]{nicklow1972lattice}%
  \BibitemOpen
  \bibfield  {author} {\bibinfo {author} {\bibfnamefont {R.}~\bibnamefont
  {Nicklow}}, \bibinfo {author} {\bibfnamefont {N.}~\bibnamefont
  {Wakabayashi}},\ and\ \bibinfo {author} {\bibfnamefont {H.}~\bibnamefont
  {Smith}},\ }\bibfield  {title} {\bibinfo {title} {Lattice dynamics of
  pyrolytic graphite},\ }\href@noop {} {\bibfield  {journal} {\bibinfo
  {journal} {Physical Review B}\ }\textbf {\bibinfo {volume} {5}},\ \bibinfo
  {pages} {4951} (\bibinfo {year} {1972})}\BibitemShut {NoStop}%
\bibitem [{\citenamefont {Zaliznyak}\ \emph {et~al.}(2017)\citenamefont
  {Zaliznyak}, \citenamefont {Savici}, \citenamefont {Garlea}, \citenamefont
  {Winn}, \citenamefont {Filges}, \citenamefont {Schneeloch}, \citenamefont
  {Tranquada}, \citenamefont {Gu}, \citenamefont {Wang},\ and\ \citenamefont
  {Petrovic}}]{zaliznyak2017polarized}%
  \BibitemOpen
  \bibfield  {author} {\bibinfo {author} {\bibfnamefont {I.~A.}\ \bibnamefont
  {Zaliznyak}}, \bibinfo {author} {\bibfnamefont {A.~T.}\ \bibnamefont
  {Savici}}, \bibinfo {author} {\bibfnamefont {V.~O.}\ \bibnamefont {Garlea}},
  \bibinfo {author} {\bibfnamefont {B.}~\bibnamefont {Winn}}, \bibinfo {author}
  {\bibfnamefont {U.}~\bibnamefont {Filges}}, \bibinfo {author} {\bibfnamefont
  {J.}~\bibnamefont {Schneeloch}}, \bibinfo {author} {\bibfnamefont {J.~M.}\
  \bibnamefont {Tranquada}}, \bibinfo {author} {\bibfnamefont {G.}~\bibnamefont
  {Gu}}, \bibinfo {author} {\bibfnamefont {A.}~\bibnamefont {Wang}},\ and\
  \bibinfo {author} {\bibfnamefont {C.}~\bibnamefont {Petrovic}},\ }\bibfield
  {title} {\bibinfo {title} {Polarized neutron scattering on hyspec: the hybrid
  spectrometer at sns},\ }in\ \href@noop {} {\emph {\bibinfo {booktitle}
  {Journal of Physics: Conference Series}}},\ Vol.\ \bibinfo {volume} {862}\
  (\bibinfo {organization} {IOP Publishing},\ \bibinfo {year} {2017})\ p.\
  \bibinfo {pages} {012030}\BibitemShut {NoStop}%
\bibitem [{\citenamefont {Savici}\ \emph {et~al.}(2017)\citenamefont {Savici},
  \citenamefont {Zaliznyak}, \citenamefont {Garlea},\ and\ \citenamefont
  {Winn}}]{savici2017data}%
  \BibitemOpen
  \bibfield  {author} {\bibinfo {author} {\bibfnamefont {A.~T.}\ \bibnamefont
  {Savici}}, \bibinfo {author} {\bibfnamefont {I.~A.}\ \bibnamefont
  {Zaliznyak}}, \bibinfo {author} {\bibfnamefont {V.~O.}\ \bibnamefont
  {Garlea}},\ and\ \bibinfo {author} {\bibfnamefont {B.}~\bibnamefont {Winn}},\
  }\bibfield  {title} {\bibinfo {title} {Data processing workflow for time of
  flight polarized neutrons inelastic measurements},\ }in\ \href@noop {} {\emph
  {\bibinfo {booktitle} {Journal of Physics: Conference Series}}},\ Vol.\
  \bibinfo {volume} {862}\ (\bibinfo {organization} {IOP Publishing},\ \bibinfo
  {year} {2017})\ p.\ \bibinfo {pages} {012023}\BibitemShut {NoStop}%
\bibitem [{\citenamefont {Laurita}(2017)}]{Laurita2017LowMagnets}%
  \BibitemOpen
  \bibfield  {author} {\bibinfo {author} {\bibfnamefont {N.~J.}\ \bibnamefont
  {Laurita}},\ }\emph {\bibinfo {title} {{Low Energy Electrodynamics of Quantum
  Magnets}}},\ \href@noop {} {Ph.D. thesis},\ \bibinfo  {school} {Johns Hopkins
  University} (\bibinfo {year} {2017})\BibitemShut {NoStop}%
\bibitem [{\citenamefont {Chauhan}\ \emph {et~al.}(2020)\citenamefont
  {Chauhan}, \citenamefont {Mahmood}, \citenamefont {Changlani}, \citenamefont
  {Koohpayeh},\ and\ \citenamefont {Armitage}}]{Chauhan2020TunableChain}%
  \BibitemOpen
  \bibfield  {author} {\bibinfo {author} {\bibfnamefont {P.}~\bibnamefont
  {Chauhan}}, \bibinfo {author} {\bibfnamefont {F.}~\bibnamefont {Mahmood}},
  \bibinfo {author} {\bibfnamefont {H.~J.}\ \bibnamefont {Changlani}}, \bibinfo
  {author} {\bibfnamefont {S.~M.}\ \bibnamefont {Koohpayeh}},\ and\ \bibinfo
  {author} {\bibfnamefont {N.~P.}\ \bibnamefont {Armitage}},\ }\bibfield
  {title} {\bibinfo {title} {{Tunable Magnon Interactions in a Ferromagnetic
  Spin-1 Chain}},\ }\bibfield  {journal} {\bibinfo  {journal} {Physical Review
  Letters}\ }\textbf {\bibinfo {volume} {124}},\ \href
  {https://doi.org/10.1103/PhysRevLett.124.037203}
  {10.1103/PhysRevLett.124.037203} (\bibinfo {year} {2020})\BibitemShut
  {NoStop}%
\bibitem [{\citenamefont {Halloran}\ \emph {et~al.}(2022)\citenamefont
  {Halloran}, \citenamefont {Wang}, \citenamefont {Li}, \citenamefont
  {Rousochatzakis}, \citenamefont {Chauhan}, \citenamefont {Stone},
  \citenamefont {Takayama}, \citenamefont {Takagi}, \citenamefont {Armitage},
  \citenamefont {Perkins},\ and\ \citenamefont {Broholm}}]{HalloranLi2IrO3}%
  \BibitemOpen
  \bibfield  {author} {\bibinfo {author} {\bibfnamefont {T.}~\bibnamefont
  {Halloran}}, \bibinfo {author} {\bibfnamefont {Y.}~\bibnamefont {Wang}},
  \bibinfo {author} {\bibfnamefont {M.}~\bibnamefont {Li}}, \bibinfo {author}
  {\bibfnamefont {I.}~\bibnamefont {Rousochatzakis}}, \bibinfo {author}
  {\bibfnamefont {P.}~\bibnamefont {Chauhan}}, \bibinfo {author} {\bibfnamefont
  {M.~B.}\ \bibnamefont {Stone}}, \bibinfo {author} {\bibfnamefont
  {T.}~\bibnamefont {Takayama}}, \bibinfo {author} {\bibfnamefont
  {H.}~\bibnamefont {Takagi}}, \bibinfo {author} {\bibfnamefont {N.~P.}\
  \bibnamefont {Armitage}}, \bibinfo {author} {\bibfnamefont {N.~B.}\
  \bibnamefont {Perkins}},\ and\ \bibinfo {author} {\bibfnamefont
  {C.}~\bibnamefont {Broholm}},\ }\bibfield  {title} {\bibinfo {title}
  {Magnetic excitations and interactions in the kitaev hyperhoneycomb iridate
  $\ensuremath{\beta}\text{\ensuremath{-}}{\mathrm{li}}_{2}{\mathrm{iro}}_{3}$},\
  }\href {https://doi.org/10.1103/PhysRevB.106.064423} {\bibfield  {journal}
  {\bibinfo  {journal} {Phys. Rev. B}\ }\textbf {\bibinfo {volume} {106}},\
  \bibinfo {pages} {064423} (\bibinfo {year} {2022})}\BibitemShut {NoStop}%
\end{thebibliography}%

\end{document}


\preprint{APS/123-QED}

\title{Supplemental information for Kitaev Magnetism in Honeycomb Iridate D$_3$LiIr$_2$O$_6$}

\date{\today}

\maketitle


\section{\label{sec:Isolation of magnetic scattering in INS Measurements} Isolation of magnetic scattering in INS Measurements}

The scattering presented in the main text was observed by three different measurements. The first was on the MACS instrument at the NCNR \cite{Rodriguez2008MACSNIST}, the second at the SEQUOIA instrument at ORNL \cite{Stone2014ASource}, and the final at the HYSPEC instrument at ORNL \cite{Stone2014ASource}. The isotopic enriched concentration of the powder sample itself was ~100\% $^2$H, ~100\%$^7$Li, 94\%$^{193}$Ir, 4\%$^{191}$Ir, and natural O as quoted from suppliers. The absorption cross-section itself per formula unit is then $\sigma_{abs}$=323.6 barns/f.u., giving a mean neutron free path length of $1/\mu_{abs} = V_0 / (2\sigma_{abs}) = 0.34 $ cm for E$_i$=25 meV neutrons that scatter elastically. $\mu_{abs}$ is the absorption length, and the factor of two comes from the number of formula units per cell, and $V_0$ is the unit cell volume of 225 $\AA^{3}$. This absorption is harshest in the MACS measurement due to the lower incident energy neutrons. 

Secondly, the isotope enriched $^2$H may in fact be imperfect, but an exact measure of this is difficult and was not provided by the supplier. The elastic scattering from the HYSPEC measurement in the $\sigma_x+\sigma_y+\sigma_z$ spin-flip channel should be purely spin-incoherent scattering, which should have the following intensity
\begin{equation}
    \int I(Q,\omega)^{SF} d\omega= \frac{\sigma_{i}}{4\pi}.
\end{equation}
Where $\sigma_{i}=6.9$ barn/f.u. assuming complete enrichment of $^2$H. However, we observe $\sigma_{i}=10.4$ barn/f.u., suggesting that about 2.5\% of the Hydrogen content is $^1$H. 

These two factors, namely absorption and significant incoherent scattering, mean that the magnetic signal is faint when compared to the effects of multiple scattering and phonon backgrounds. The extraction of the purely magnetic part is detailed step by step for each respective measurement within the subsections below. 

\subsection{SEQUOIA Measurement Details}

The primary measurement presented in the text was measured on the SEQUOIA instrument, as it has the greatest statistics and lowest absorption thereby giving the best access to the magnetic scattering. The measurement itself was performed using a CCR style cryostat with 3.6(1) g of isotope enriched sample. The sample was loaded into a custom annular aluminum can of 20 mm outer diameter and 19 mm inner diameter with height 4 cm and was prepared by solid state synthesis methods as described in Ref. \cite{Kitagawa2018ALattice} with triply isotope enriched $^2$D, $^{193}$Ir, and $^{7}$Li to mitigate absorption and incoherent scattering. A vanadium standard was measured using each experimental configurations to normalize both experiments to barn/sr/eV/mol Ir. Annular angle-dependent absorption was corrected by a Monte-Carlo method \cite{Schmitt1998AbsorptionDiffraction}. Two sets of measurements were taken at each temperature, one with E$_i$=30 meV neutrons and the 300 Hz high resolution chopper configuration and the other with the E$_i$=60 meV 420 Hz high flux chopper configuration. Counting times are summarized below.
\begin{table}
\begin{tabular}{|p{0.2\columnwidth}|p{0.35\columnwidth}|p{0.15\columnwidth}|p{0.15\columnwidth}|}
 \hline
 E$_i$ (meV) & Chopper & T (K) & Time (h)  \\
 \hline
 \hline
30  & High Res 300 Hz  & 4.5(1) & 22 \\
\hline
30  & High Res 300 Hz  & 100.0(1) & 14 \\
\hline
30  & High Res 300 Hz  & 200.0(1) & 12 \\
\hline
60  & High Flux 420 Hz  & 4.5(1) & 22 \\
\hline
60 & High Flux 420 Hz  & 100.0(1) & 10 \\
\hline
60& High Flux 420 Hz & 200.0(1) & 10 \\
\hline
\end{tabular}
\caption{Table of experimental configuration for SEQUOIA measurements on D$_3$LiIr$_2$O$_6$. In this table, one hour of counting corresponds to 4.2 C of proton charge from the spallation source.}
\label{table:SEQ_configs}
\end{table}

\begin{figure}
    \centering
    \includegraphics[width=1.0\columnwidth]{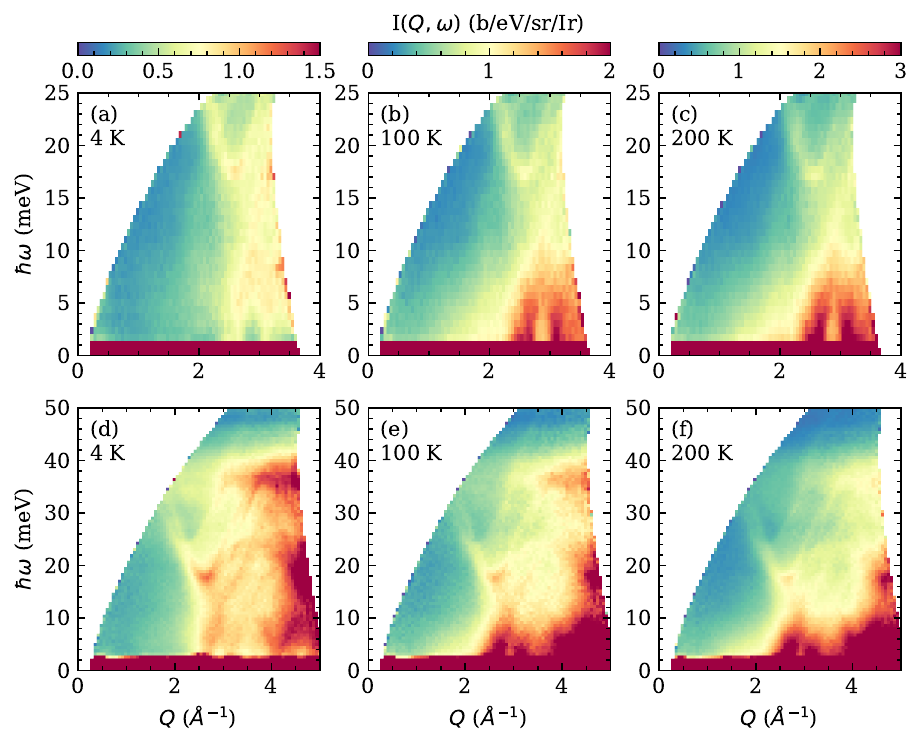}
    \caption{Plots of absorption-corrected scattering for all six experimental configurations from the SEQUOIA experiment for D$_3$LiIr$_2$O$_6$. The E$_i$=30 meV is presented in (a-c), and E$_i$=60 meV in (d-f). Intensity scale is constant within each temperature.}
    \label{fig:dlio_seq_configs}
\end{figure}
The treatment of the data is summarized by the following steps.

\begin{enumerate}
    \item Begin with raw counts from measurements of E$_i$=30 meV, 60 meV, and $T$=2 K, 100 K, and 200 K for each. 
    \item A vanadium standard is used to normalize the scattering to absolute units. 
    \item Energy-dependent absorption is corrected by means of a Monte-Carlo method built into the Mantid suite. 
    \item All scattering is corrected using a temperature-dependent Debye-Waller factor using the elastic scattering.
    \item Phonon scattering is subtracted from the low temperature measurements by assuming a Bose-Einstein population of phonons as well as assuming no magnetic scattering in the high temperature measurements. 
    \item Remove effects of multiple scattering using a particular form that assume spin-incoherent elastic scattering and phonon events.
    \item The remaining scattering is then analyzed using a factorization approach that assumes that the $Q$ and $\omega$ dependence of the scattering is independent, which is presented in the main text.
\end{enumerate}

The result of steps 1-3 for each measurement is shown in Fig. \ref{fig:dlio_seq_configs}(a-f). It is immediately obvious that both configurations at T=4 K are dominated by phonon scattering and what seems to be a constant background that originates from the nuclear spin-incoherent scattering of D/H. Very little difference is seen in either configuration between the T=100 K and T=200 K measurements. Magnetic scattering cannot be directly resolved from these measurements. Energy dependent absorption is handed for each neutron event using the AnnularRingAbsorption Monte Carlo method built into the Mantid software. The precise dimensions of our annular can used for this were an outer diameter of 20 mm, inner diameter of 19 mm, and height of 38 mm (1mm annulus of sample). 

In general, neutron scattering cross sections include the Debye-Waller factor which is an overall prefactor to intensity that goes as $e^{-\langle u^2 \rangle Q^2}$ where $\langle u^2 \rangle$ is the mean-squared ion displacement. Of course, this is not a constant for every ion in the unit cell but as a first-order approximation we take it to be one value for all ions and scattering processes. The elastic scattering should then go as 

\begin{equation}
    I(Q,\omega=0) = e^{-\langle u^2 \rangle Q^2}(\sigma_{bragg}(Q) + \sigma_{i}).
\end{equation}

\begin{figure}
    \centering
    \includegraphics[width=1\columnwidth]{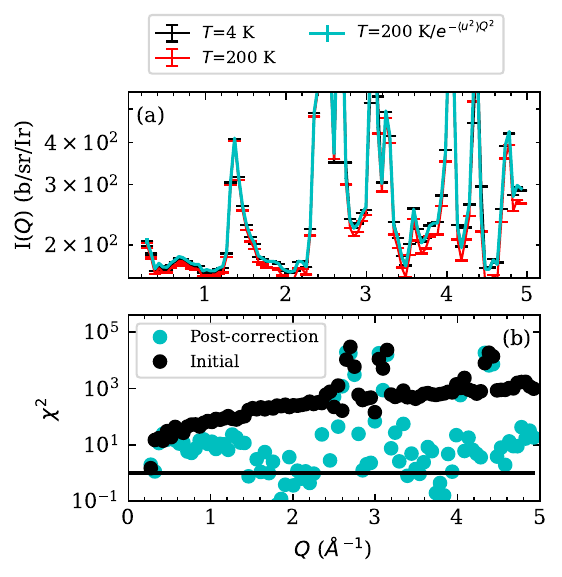}
    \caption{(a) Elastic cuts of E$_i$=60 meV Elastic scattering for the $T$=4 K and $T$=200 K SEQUOIA measurements averaged from $\hbar\omega=\pm 1$ meV. (b) Deviation between the two measurements as defined by $[\sum (I_{4K}(Q)-I_{200K}(Q))^2/(\delta I_{4K}(Q)^2 + \delta I_{200K}(Q)^2)]$. The black points represent before correcting for the Debye-Waller factor and the cyan points after. The solid black line is fixed at $\chi^2=1$. Error bars in this and all following figures represent one standard deviation.}
    \label{fig:DW_fig}
\end{figure}

The mean-square displacement $\langle u^2 \rangle$ is temperature-dependent. If one assumes that at $T$=4 K $\langle u^2 \rangle$=0, the difference between elastic cuts as a function of $Q$ may be used to extract the Debye-Waller factor for each elevated temperature (100 K and 200 K). The higher incident energy configuration has access to a wider range of $Q$, so cuts of these measurements averaged from $\hbar\omega=\pm 1$ meV are used for this purpose. The extracted values of $\langle u^2 \rangle$ are  $\langle u^2 \rangle_{4K}=0$ (assumed), $\langle u^2 \rangle_{100K}$=1.1(1) m$\AA^2$, and $\langle u^2 \rangle_{200K}$=3.3(1) m$\AA^2$. An example of the elastic cuts before and after applying this correction is shown in Fig. \ref{fig:DW_fig}. 

Because of the relatively large incoherent scattering cross section of D ($\sigma_{i}$=4.04 b) and the very large incohenent scattering from H due to imperfect enrichment ($\sigma_{i}^{H}=25.274$ b), we attribute the presence of a large flat background across the entire measurement to a multiple scattering event between a strong elastic incoherent process and a weaker inelastic scattering process. This is most simply taken to be a phonon. To model such an event, we begin with the individual cross sections for incoherent elastic and incoherent inelastic phonon scattering

\begin{equation}
    \frac{d\sigma}{d\Omega}_{el\ inc} = \frac{\sigma_{inc}}{4\pi}Ne^{-2W},
    \label{eq:el_inc}
\end{equation}
\begin{equation}
    \frac{d^2\sigma}{d\Omega d\omega}_{inel\ inc}=\frac{k_f}{k_i}\frac{\sigma_{inc}}{4\pi}\frac{3N}{2M} \frac{Z(\omega)}{\omega} \frac{e^{\beta \omega}}{e^{\beta \omega} -1}Q^2 e^{-2W}.
    \label{eq:inel_inc}
\end{equation}

Here, $\sigma_{inc}$ is the incoherent cross section per formula unit and $Z(\omega)$ is related to the phonon density of states. We also take the Debye-Waller factor $e^{-2W}$ to be one. This is the same as the treatment used in Ref. \cite{Hong2006NeutronParamagnet}. We now assume a thin sample with a simplified form for the phonon cross section of I$_p(Q,\omega)$=$f(\omega)Q^2$ with all the constants and energy dependence of Eq. \ref{eq:inel_inc} being absorbed in $f(\omega)$. There are three scattering vectors involved in this process. The first $\textbf{k}_i$ is fixed by the incoming neutron configuration and beam direction. The second $\textbf{k}_i'$ is the intermediate scattering vector after the first incoherent elastic event. The final is $\textbf{k}_f$, which is measured at the detector. Given the initial fixed $\textbf{k}_i$ and that ever observed $\textbf{Q}=\textbf{k}_i - \textbf{k}_f$, we may find the effect of multiple scattering by integrating over all allowed $\textbf{k}_i'$. Because we have a powder sample, this will be a spherical average. So, the minimum allowed value for $k_i'$ is given by $|\textbf{k}_i'|=|k_i - k_f|$ and the maximum is given by $|k_i+k_f|$. We now write
\begin{eqnarray}
    I(Q,\omega)=T I_0(Q,\omega) + \nonumber\\(1-T)f(\omega)\int d\Omega \int_{|k_i - k_f|}^{|k_i+k_f|}(|\textbf{k}_i' - \textbf{k}_f|)^2 d\textbf{k}_i'
    \label{eq:mult_scatt_1}
\end{eqnarray}
Here $I_0(Q,\omega)$ represents all single event scattering, $T$ represents the fraction of neutrons that only experience a single scattering event, $C$ contains the constants in Eqs. \ref{eq:el_inc} and \ref{eq:inel_inc}, and the $'$ superscript denotes the intermediate scattering process. The integral is readily evaluated and we find that the overall scattering is given by 
\begin{equation}
    I'(Q,\omega) = f(\omega)\bigg( TQ^2 + (1-T)(k_i^2+ k_f^2)\bigg).
    \label{eq:final_mc_eq}
\end{equation}
We may rewrite this in the easily linearized form 
\begin{equation}
    I'(Q,\omega) = f(\omega)(k_i^2+ k_f^2)\bigg( \frac{TQ^2}{(k_i^2+ k_f^2)} + (1-T)\bigg).
    \label{eq:final_mc_eq_linear}
\end{equation}

\begin{figure}
    \centering
    \includegraphics[width=\columnwidth]{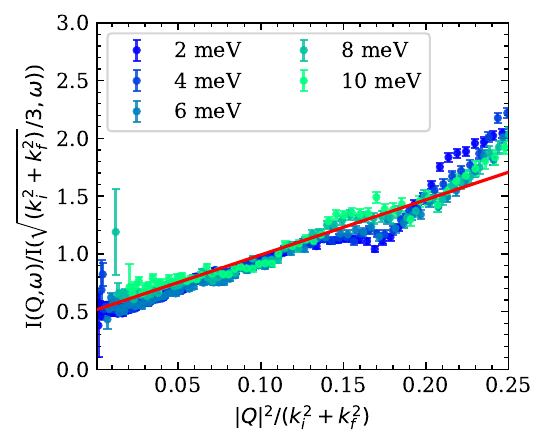}
    \caption{Collapse of E$_i$=30 meV T=200 K measurement to form of Eq. \ref{eq:final_mc_eq_linear}. The slope of the collapse provides the quantity of multiple scattering in the system. Cuts have a total energy range of 1 meV, meaning that they average over $\pm$ 0.5 meV of the value indicated in the legend.}
    \label{fig:mult_scatter_collapse}
\end{figure}
An observation to be made about this form is its strong dependence on $k_i$ and $k_f$. For the E$_i$=30 meV measurement with energy transfer $\hbar\omega=3$ meV, ($k_i^2$+ $k_f^2$)=27.5 $\AA^{-2}$, whereas for the MACS measurement with E$_f=$5.0 meV and energy transfer $\hbar\omega=3$ meV, ($k_i^2 +k_f^2$)=6.3 $\AA^{-2}$, meaning that this effect would be suppressed by nearly an order of magnitude simply due to kinematics. We assume the single-event scattering fraction $T$ to be constant across measurements. The value of this parameter is found using the linearized form of Eq. \ref{eq:final_mc_eq_linear}, which is visually represented in Fig. \ref{fig:mult_scatter_collapse}. The assumption that the functional form for phonons is simply $I_p(Q,\omega)=f(\omega)Q^2$ breaks down significantly at high Q and energy due to anharmonicity which is most clearly visible in Fig. \ref{fig:dlio_seq_configs}(d), but within the $Q$ and energy range of our interest for magnetic scattering it works quite well and from which we extract an overall single-event scattering fraction of $T=0.90(1)$. 

\begin{figure}
    \centering
    \includegraphics[width=1.0\columnwidth]{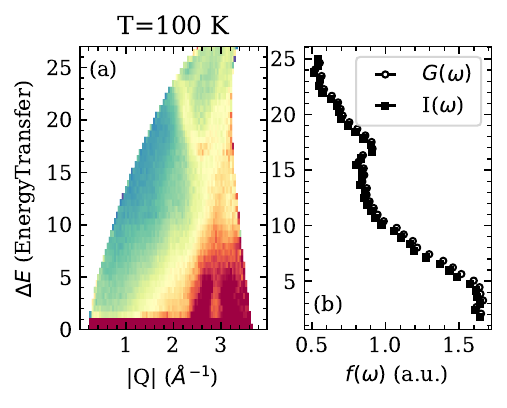}
    \caption{Example of extraction of $f(w)$ quantity for calculation of multiple scattering background. Scattering shown is from the E$_i$=30 meV T=100 K measurement.}
    \label{fig:factorizatoin_fw}
\end{figure}

We now require a form of $f(\omega)$ from which to compute $I'(Q,\omega)$. Based on the magnetic form factor of Ir$^{4+}$ we assume that the extreme values of $Q$ available in each measurement represent the purely non-magnetic contributions to the scattering. These regions in $Q$ are from $Q$=2.5 $\AA^{-1}$ to $Q$=3.0 $\AA^{-1}$ for the E$_i$=30 meV measurements and $Q$=3.5 $\AA^{-1}$ to 4.0 $\AA^{-1}$ for the E$_i$=60 meV measurements. Within these widows, both a simple cut along the energy transfer dimension or a factorization approach described later works equally well. For completeness we use the factorization. This particular model of phonon and multiple scattering assumes a form of I$_p(Q,\omega)$=$f(\omega)Q^2$ for single-event scattering. It is then sensible that an attempt to subtract $I'(Q,\omega)$ will fail in $Q-\omega$ regimes where this assumption does not hold, many of which can be seen in Fig. \ref{fig:dlio_seq_configs}. However, we may safely assume that these deviations from the assumed form of $I'(Q,\omega)$ are dominated by single-event phonon scattering. So, the total scattering intensity may now be written as 

\begin{equation}
    I_{T}(Q,\omega) = I^{mag}_{T}(Q,\omega)+I'_{T}(Q,\omega)+\delta I_{T}^{ph}(Q,\omega).
    \label{eq:total_scattering_anharmonic}
\end{equation}

The subscript $T$ denotes the temperature, $I^{mag}$ is the magnetic contribution, and $\delta I^{ph}$ is the anharmonic phonon contribution dominated by single-phonon scattering. Equipped with $f^{T}(\omega)$, $I'_T(Q,\omega)$ is readily calculated for each configuration. The dominant single-event phonon contributions are now subtracted using a standard Bose-Einstein form, and assuming no magnetic scattering for high temperatures this leaves 
\begin{align}
    &\bar{I}(Q,\omega) = I^{T_L}(Q,\omega) - \frac{1-e^{-\beta_H\omega}}{1-e^{-\beta_L\omega}} I^{T_H}(Q,\omega) \\
    \label{eq:Ibar_L}
    &\bar{I}(Q,\omega) = I^{mag}_{T_L} + I'_{T_L}(Q,\omega) -  \frac{1-e^{-\beta_H\omega}}{1-e^{-\beta_L\omega}} I'_{T_H}(Q,\omega).
\end{align}

\begin{figure}
    \centering
    \includegraphics[width=1\columnwidth]{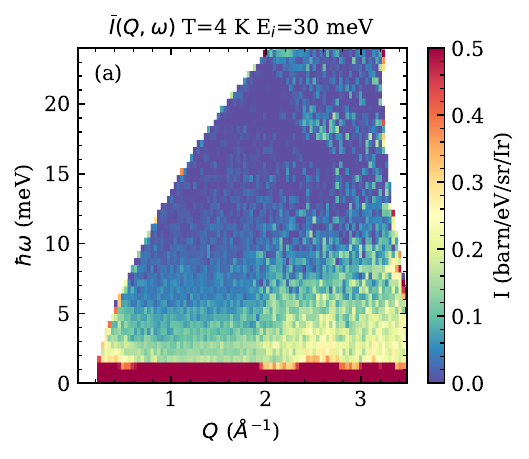}
    \includegraphics[width=1\columnwidth]{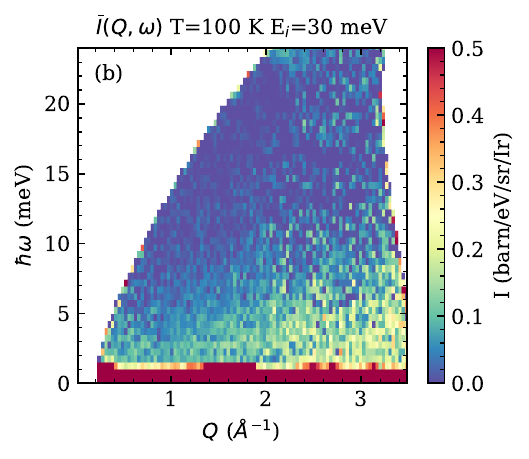}
    \caption{Remaining scattering after subtracting high temperature measurements. T=100 K was used as a background for the T=4 K measurement in (a), and the T=200 K measurement was used as a background for the T=100 K measurement in (b). }
    \label{fig:Ibar_4K_100K_30meV}
\end{figure}

The intensity in the quantity $\bar{I}$ for each temperature still retains a nonmagnetic background from multiple scattering, which may now be defined as 

\begin{equation}
    I_{bkg}(Q,\omega) =  I'_{T_L}(Q,\omega) -  \frac{1-e^{-\beta_H\omega}}{1-e^{-\beta_L\omega}} I'_{T_H}(Q,\omega),
    \label{eq:I_bkg_total}
\end{equation}

\begin{figure}
    \centering
    \includegraphics[width=1.0\columnwidth]{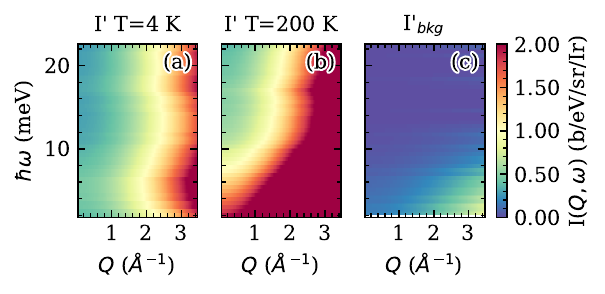}
    \caption{(a,b) Calculated multiple scattering backgrounds $I'(Q,\omega)$ for $T=4$ K and $T=100$ K E$_i$=30 meV measurements using form of Eq. \ref{eq:final_mc_eq_linear}. (c) Resulting background after Bose-Einstein subtraction of phonons as defined by Eq. \ref{eq:I_bkg_total}.}
    \label{fig:nonmag_bkg}
\end{figure}

which can be explicitly calculated as depicted in Fig \ref{fig:nonmag_bkg}. An overall normalization factor $A$ is applied such that the resulting magnetic scattering at high $Q$ and $\omega$ fluctuates around zero. Finally, the scattering presented in the text is given by 

\begin{equation}
    I_{mag}(Q,\omega) = \bar{I}^{T_L}(Q,\omega) - AI_{bkg}(Q,\omega)
    \label{eq:final_sub}
\end{equation}

\begin{figure}
    \centering
    \includegraphics[width=1.0\columnwidth]{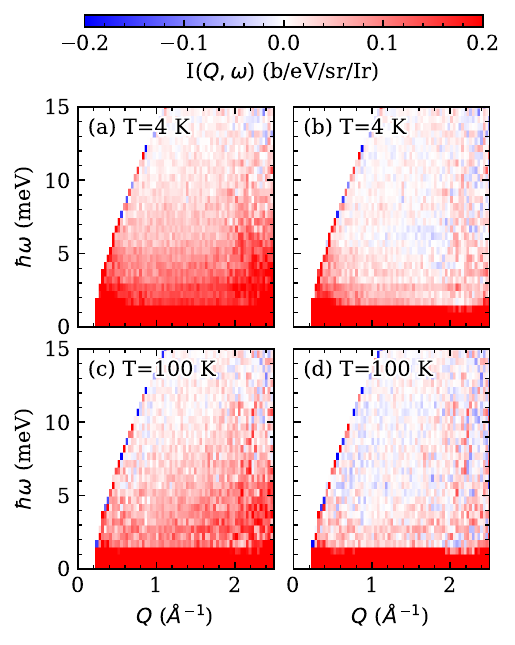}
    \caption{Demonstration of the effect of the removal of multiple scattering background for E$_i$=30 meV measurements. (a) and (c) show $\bar{I}(Q,\omega)$ as defined in Eq. \ref{eq:Ibar_L} and (b) and (d) show the result of the subtraction in Eq. \ref{eq:final_sub}. The temperature dependence of the magnetic scattering is made clear by a diverging color scale, showing an almost perfect subtraction in (d) but a clear remainder signal in (b).}
    \label{fig:.}
\end{figure}

\subsection{MACS Measurement Details}

The measurement performed on the MACS instrument used a 2.0 g sample of enriched $^{2}$D$_3^{7}$Li$^{193}$Ir$_2$O$_6$. The four different experimental configurations are summarized in Table \ref{table:MACS_configs}.  The analysis of the scattering from the MACS experiment may be summarized in the following:

\begin{enumerate}
    \item Scattering was measured for T=1.7 K and T=55 K for both E$_f$=3.7 meV and E$_f$=5.0 meV configurations. Matching empty can background measurements were taken for both E$_f$ configurations.
    \item Empty can backgrounds were subtracted with an appropriate self-shielding factor by using the available Aluminum elastic Bragg peaks as a reference point. 
    \item Energy dependent absorption was corrected by a numerical method using DAVE~\cite{azuah2009dave}. 
    \item Scattering was normalized to absolute units using the scattering intensity of all available Bragg peaks. 
    \item Remove phonon scattering by the same temperature subtraction as the SEQUOIA measurement.
    \item Remnant elastic scattering exits in the inelastic channel of these measurements due to phonon scattering from the monochromator. This effect is removed by tracking (001) Bragg peak intensity as a function of energy. The remainder is assumed to be the purely magnetic contribution.
\end{enumerate}

For the normalization, the sum of all available Bragg peak intensities was compared to the expected integrated intensity based on nuclear structure factors. The elastic intensity as a function of $Q$ originating from Bragg scattering should go as 
\begin{table}
\begin{tabular}{|p{0.3\columnwidth}|p{0.2\columnwidth}|p{0.4\columnwidth}|}
 \hline
 E$_f$ (meV) & T (K) & Counting Time (h)  \\
 \hline
 \hline
3.7  &  1.72(1) & 9.9 \\
\hline
3.7  &  55.0(1) & 9.9 \\
\hline
5.0  &  1.72(1) & 19.1 \\
\hline
5.0  &  55.0(1) & 20.8 \\
\hline
\end{tabular}
\caption{Table of experimental configuration for MACS measurements on D$_3$LiIr$_2$O$_6$.}
\label{table:MACS_configs}
\end{table}

\begin{equation}
    I(Q,\omega) = {\cal A} \frac{d^2 \sigma}{d\Omega d\omega}
    \label{eq:I_general}
\end{equation}
where ${\cal A}$ is the normalization factor. The coherent nuclear scattering cross section is readily evaluated and provides the following form for ${\cal A}$

\begin{equation}
    {\cal A}=N\frac{4\pi\int I(Q,\omega)Q^2 dQ d\omega}{M|F_{HKL}|^2\frac{(2\pi^3)}{V_0}}
    \label{eq:Norm_factor}
\end{equation}
The integral represents the measured Bragg peak intensity, $|F_{HKL}|^2$ is the nuclear structure factor, $M$ is the multiplicity of the Bragg peak, $N$ is number of formula units of sample, and $V_0$ is the unit cell volume. In the case of MACS, we assume that the integrated intensity along the energy axis is given by the multiplication of the elastic line by the elastic resolution FWHM.

\begin{figure}
    \centering
    \includegraphics[width=1.0\columnwidth]{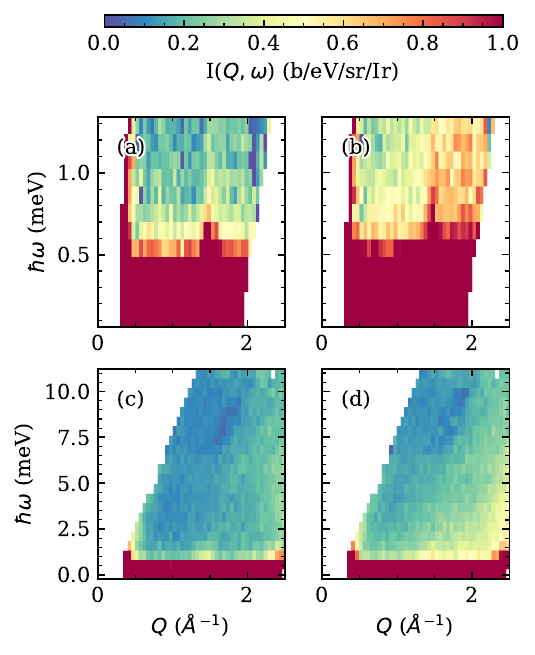}
    \caption{Normalized scattering intensity from the MACS instrument for the four main experimental configurations.  E$_f$=3.7 meV $T$=1.72(1) K is shown in (a), E$_f$=3.7 meV $T$=55.0(1) K is shown in (b), E$_f$=5.0 meV $T$=1.72(1) K is shown in (c), and E$_f$=5.0 meV $T$=55.0(1) K is shown in (d). Scattering has been corrected for absorption and a background measurement has been subtracted.}
    \label{fig:macs_scatter_configs}
\end{figure}

After empty can subtractions, absorption corrections, and normalization, the four measurements are summarized in Fig. \ref{fig:macs_scatter_configs}(a-d). 

One immediate observation in the E$_f$=3.7 meV measurements is what looks like the (001) Bragg peak extending far into the inelastic channel. While at first glance this seems to be associated with the instrumental resolution,  the expected FWHM in energy transfer of the MACS instrument in this configuration on the order of $\Delta\hbar\omega=$0.1 meV. This effect, which may be summarized as the elastic line leaking into the inelastic channel, has been seen before both on MACS and other triple-axis spectrometers and ultimately originates from allowed inelastic scattering processes from the monochromator. For the inelastic channel, we consider both of the following processes to be equally likely:

\begin{itemize}
    \item Bragg (002) reflection from graphite monochromator (strong process), followed by inelastic magnetic scattering at the sample (weak process). 
    \item Acoustic phonon scattering originating from the (002) Bragg position at the graphite monochromator (weak process), followed by elastic scattering at the sample position (strong process). 
\end{itemize}

Because elastic cross sections are order of magnitude higher than inelastic, we these to be strong and inelastic scattering such as phonon or magnetic scattering to be weak. Before discussing the precise mechanism of this effect, we may describe the scattering using the following functional form:

\begin{equation}
    I(Q,\omega)= A(\omega)I_{el}(Q) + B(\omega)I_{mag}(Q) + I_{ph}(Q,\omega).
\end{equation}
The lineshape $I_{el}(Q)$ is the elastic lineshape that persists to finite energy transfers, $I_{mag}(Q)$ is $Q$-dependent form of the magnetic scattering, and $I_{ph}(Q,\omega)$ is the dominant contribution to the background which is phonon scattering. Performing the same Bose-Einstein subtraction of phonons as before produces the following:
\begin{align}
    &\bar{I}(Q,\omega) = I_{L}(Q,\omega)-\frac{1-e^{-\beta_H \omega}}{1-e^{-\beta_L \omega}}I_{H}(Q,\omega)\\
    &\bar{I}(Q,\omega) = A(\omega)(1-\frac{1-e^{-\beta_H \omega}}{1-e^{-\beta_L \omega}})I_{el}(Q) + B(\omega)I_{mag}(Q).
    \label{eq:ibar_macs}
\end{align}

The subscripts $H$ and $L$ indicate high and low temperature measurements respectively. Because the Bose-Einstein factor has no $Q$ dependence and the form of I$_{mag}(Q,\omega)$ may be assumed to be of form of nearest-neighbor correlated spins, we may write Eq. \ref{eq:ibar_macs} as

\begin{equation}
    \bar{I}(Q,\omega)=C(\omega)I_{el}(Q)+B(\omega)(1+\frac{\sin(Qd)}{Qd}).
    \label{eq:macs_elscatter_model}
\end{equation}

\begin{figure}
    \centering
    \includegraphics[width=1.0\columnwidth]{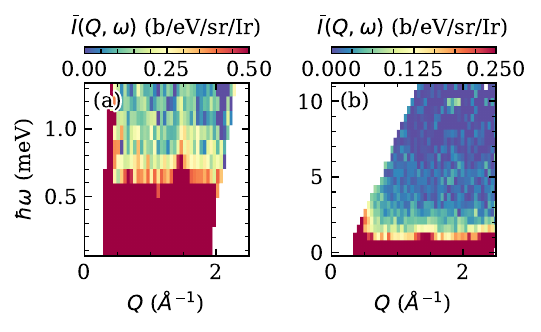}
    \caption{Scattering from the MACS experiment after subtraction of high temperature data as defined by Eq. \ref{eq:ibar_macs} from the E$_f$=5.0 meV (a) and E$_f$=3.7 (b) configurations.}
    \label{fig:Ibar_macs}
\end{figure}

The nearest neighbor distance $d$ is 3.54 $\AA$, and the parameters $C(\omega)$ and $B(\omega)$ are constants allowed to fit at each energy transfer. The quantity $\bar{I}$ is presented in Fig. \ref{fig:Ibar_macs}

The remaining scattering of elastic origin is completely dominant in the E$_f$=3.7 meV $\bar{I}$ (Fig. \ref{fig:Ibar_macs}(b)) and the elastic Bragg peak is seen in the E$_f$=5.0 meV measurement clearly up to energy transfer of $\hbar\omega$=2 meV. To understand the origin of this peculiar background, we must first consider the scattering condition that allows for a neutron to scatter at the correct $2\theta$ to travel from the neutron source and scatter from the monochromator towards the sample

\begin{equation}
    2\theta_{mono} = \cos^{-1}(k_{i}^2 + k_f^2 - |\tau_{002}|^2 / (2k_{i}k_f)).
    \label{eq:twotheta_mono}
\end{equation}

\begin{figure}
    \centering
    \includegraphics[width=1\columnwidth]{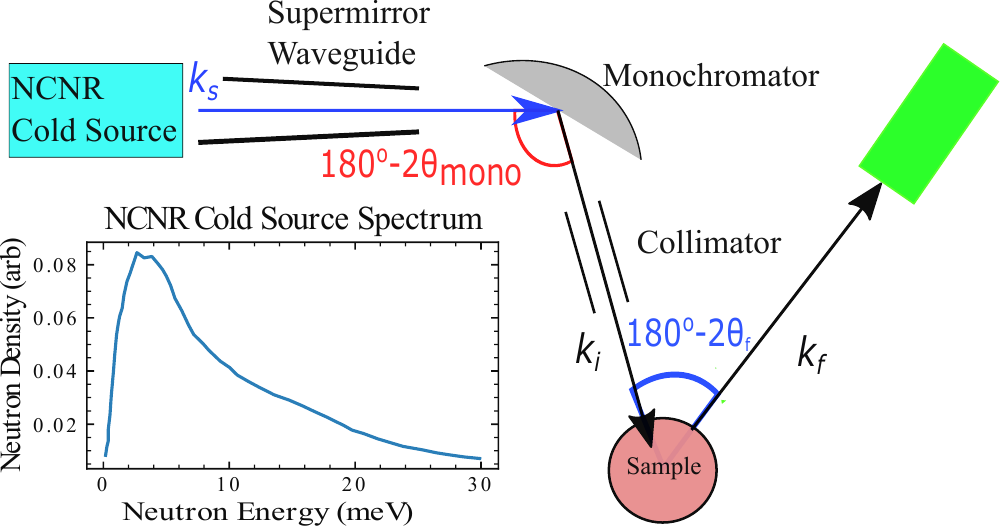}
    \caption{Sketch showing generic scattering process for a triple-axis spectrometer, which defines the value of 2$\theta_{mono}$ in Eq. \ref{eq:twotheta_mono}. The inset shows the density of incident neutron energies from the cold source. The green box is a simplified version of the double focusing analyzer and detector geometry.}
    \label{fig:mono_sketch}
\end{figure}

Eq. \ref{eq:twotheta_mono} is used to calculate the scattering angle of the Bragg reflection from the (002) reflection of pyrolytic graphite which is the origin of $|\tau_{002}|^2$. Here, $k_i$ is expected wavevector for elastically scattered neutrons at the monochromator and $k_f$ is defined by the analyzers, and one may readily compute the scattering angle $2\theta_{mono}$. The value of $2\theta_{mono}$ is fixed, and depicted in Fig. \ref{fig:mono_sketch}. If we loosen this condition such that arbitrary wavevectors from the cold source are incident on the monochromator, we may solved for the allowed incoming wavevector as a function of the momentum transfer at the monochromator
\begin{equation}
    Q_{mono}^2 = k_s^2 + k_f^2 - 2k_sk_fcos(2\theta_{mono}).
    \label{eq:reflect_condition}
\end{equation}

\begin{figure}
    \centering
    \includegraphics[width=1.0\columnwidth]{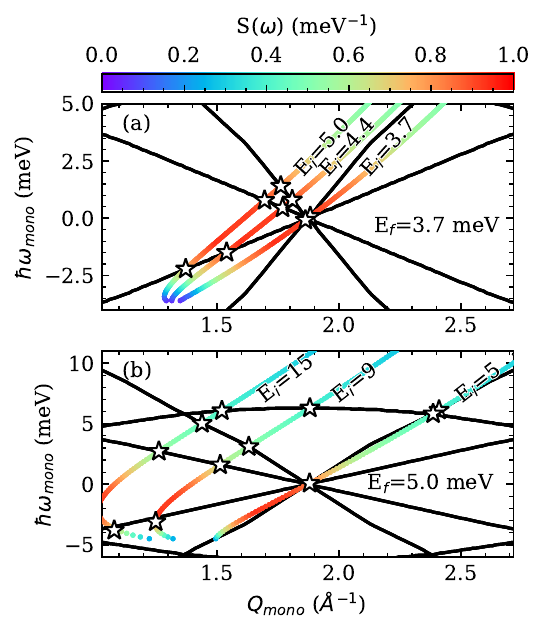}
    \caption{Depiction of allowed scattering vectors on the MACS instrument considering inelastic scattering at the monochromator. The $Q$ direction is along the (00$L$) axis of pyrolytic graphite, and the color map shows potential scattering vectors that satisfy the 2$\theta$ condition to scatter from the monochromator to the sample. Solid black lines show phonon dispersion of pyrolytic graphite along the (00$L$) direction reproduced from Ref.~\cite{nicklow1972lattice}. Black squares represent the elastic scattering configuration. }
    \label{fig:MACS_allowed_scatter}
\end{figure}
The wavevector $k_s$ is an arbitrary wavevector for incident neutrons, and the momentum transfer at the monochromator is no longer fixed to be $|\tau_{002}|^2$ but instead an arbitrary value $Q_{mono}$. Of course, the elastic scattering condition with $Q=|\tau_{002}|^2$ will satisfy this condition, but using the distribution of neutrons incident on the monochromator from the cold source we find a large map of allowed scattering in $Q-\omega$ space. To further refine this, we may weight this map by the spectral density of neutrons from the cold source which will be referred to as $S(\omega)$, which is reproduced in the inset of Fig. \ref{fig:mono_sketch}.  

The color map in Fig. \ref{fig:MACS_allowed_scatter} depicts all allowed scattering paths that satisfy the scattering condition in Eq. \ref{eq:reflect_condition}. 2$\theta_{mono}$ is fixed by the E$_i$ setting, $k_f$ is fixed by the E$_f$ setting, so using Eq. \ref{fig:MACS_allowed_scatter} one may calculate the $k_S$ wavevector for incoming neutrons for any momentum transfer at the monochromator $Q$. The result is the color map, upon which the low energy dispersion of pyrolytic graphite along the (00$L$) direction has been plotted in black lines. The intercept of the color plot and the black lines represents a direct allowed path for inelastic scattering from the monochromator to reach the sample for these particular E$_f$ configurations. The spectral weight from the cold source and the $1/\omega$ contribution to the neutron phonon cross section both contribute to the absence of this effect at higher energy transfers. 

\begin{figure}
    \centering
    \includegraphics[width=1.0\columnwidth]{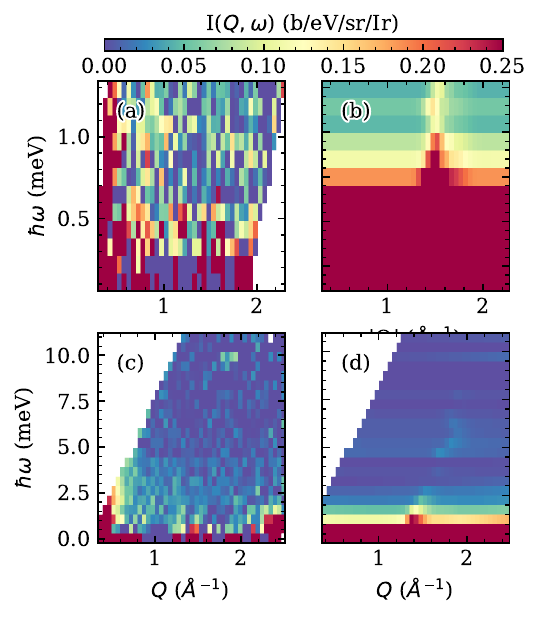}
    \caption{Overall extracted magnetic intensity from the MACS measurement in both the E$_f$=3.7 meV (a) configurations and the E$_f$=5.0 meV (c) configurations. The extracted elastic backgrounds present in $\bar{I}(Q,\omega)$ are shown in (b) and (d) for the E$_f$=3.7 and E$_f$=5.0 measurements respectively.}
    \label{fig:macs_el_bkg}
\end{figure}

\subsection{Polarized HYSPEC Measurement Details}

Polarized inelastic neutron scattering was done at the HYSPEC instrument at ORNL. The measurement was performed with incident energy E$_i$=20 meV with the intermediate flux configuration. The sample used in the SEQUOIA experiment was resused but reloaded into a 7 mm outer diameter 6 mm inner diameter annular can. The HYSPEC measurement was performed in the 3D polarized mode in the spin-flip and non spin-flip settings~\cite{zaliznyak2017polarized}, as summarized in Table \ref{table:HYS_configs}. Annular absorption corrections were applied to all configurations. The challenge in the analysis of these data originate from extremely low statistics with a counting rate of nearly 10 times less than the SEQUOIA measurement and the sensitivity of the detector normalization to the sample geometry. This sensitivity originates from the visibility of the sample at each detector through the unique supermirror polarizing array. The analysis of these data utilizes a modified version of a previously published algorithm~\cite{savici2017data}, and may be summarized by the following: 

\begin{enumerate}
    \item The scattering was sorted into the six respective cross sections for 3D polarized scattering, $\sigma_{x,y,z}^{SF,NSF}$.
    \item Absorption was corrected using a Monte-Carlo method in the MANTID software.
    \item A flipping ratio correction is performed such that coherent elastic scattering from the sample vanishes in the spin-flip channel. The (001) Bragg peak was used as a guide for this. 
    \item The elastic scattering from the spin-flip total scattering is assumed to be purely spin-incoherent sample scattering (no magnetic scattering). This is then used as the detector normalization.
    \item This normalization and flipping ratio correction is then verified by the resulting six elastic scattering lines. 
    \item The scattering is normalized to absolute units using the accessible Bragg peak intensities. 
    \item The true magnetic scattering and other relevant quantities of interest may then be extracted. 
\end{enumerate}

\begin{table}
\begin{tabular}{|p{0.15\columnwidth}|p{0.35\columnwidth}|p{0.2\columnwidth}|p{0.15\columnwidth}|}
 \hline
 E$_i$ (meV) & Chopper & Polarization & Time (h)  \\
 \hline
 \hline
20  & Intermediate 240 Hz  & $\sigma_x^{SF}$ & 30 \\
\hline
20  & Intermediate 240 Hz  & $\sigma_y^{SF}$& 30 \\
\hline
20  & Intermediate 240 Hz  &$\sigma_z^{SF}$ & 66 \\
\hline
20  & Intermediate 240 Hz  & $\sigma_x^{NSF}$ & 12 \\
\hline
20 & Intermediate 240 Hz  & $\sigma_y^{NSF}$ & 12 \\
\hline
20& Intermediate 240 Hz & $\sigma_z^{NSF}$ & 8 \\
\hline
\end{tabular}
\caption{Table of experimental configuration for HYSPEC measurements on D$_3$LiIr$_2$O$_6$. In this table, one hour of counting corresponds to 5 Coulombs of proton charge from the spallation source. }
\label{table:HYS_configs}
\end{table}
In general, the correction to polarized neutron scattering due to the contribution from the imperfect flipping ratio of the supermirror polarizing array is given by 

\begin{align}
    I_{SF_c}=\frac{F}{F-1}I_{SF} - \frac{1}{F-1}I_{NSF}\nonumber \\
    I_{NSF_c}=\frac{F}{F-1}I_{NSF}-\frac{1}{F-1}I_{SF}.
\end{align}

Using these corrections, one would expect no elastic Bragg scattering in the spin-flip channel with the correct value of $F$. To find $F$ in the $x,y,$ and $z$ spin-flip configurations, $F$ was fit such that there was no observable Bragg scattering in the spin-flip channel along the elastic line. For this purpose, cuts averaged along energy transfers of $\hbar\omega=\pm1$ meV were used. From this, the flipping ratio is taken to be $F=13$ for all polarization directions.

\begin{figure}
    \centering
    \includegraphics[width=1.0\columnwidth]{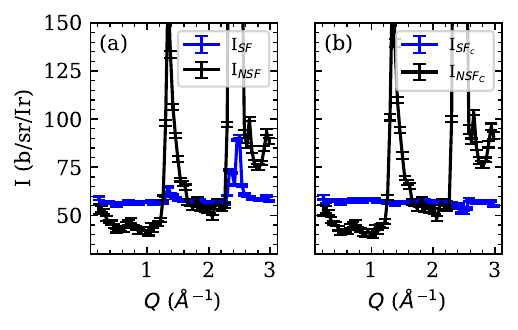}
    \caption{Elastic cuts from HYSPEC experiment averaged in $\hbar\omega=\pm1$ meV. The $\sigma_z$ cross section is shown, with the direct (a) and flipping ratio corrected (b) spin-flip and non spin-flip intensities are pictured in blue and black respectively. . }
    \label{fig:hys_flip_ratio}
\end{figure}

\begin{figure}
    \centering
    \includegraphics[width=1\columnwidth]{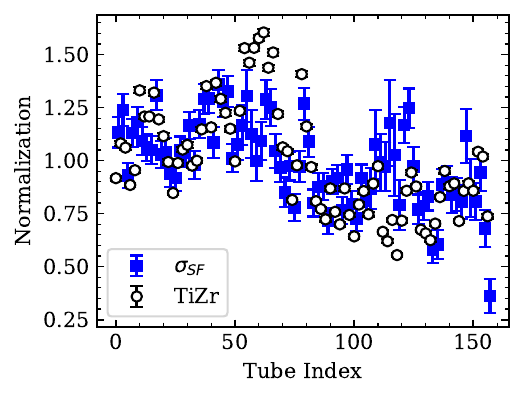}
    \caption{Normalization extracted from the elastic ($\hbar\omega=\pm1$ meV) total spin-flip scattering in the HYSPEC measurement. The result in blue shows overall good agreement with the TiZr standard, shown in black. Only one half of points are visualized.}
    \label{fig:HYS_norm}
\end{figure}

A specific problem associated with polarized scattering on the HYSPEC instrument is that the polarizing supermirror array acts as a radial collimator. While this is normally excellent for removing background from the sample environment, it also means that each detector's visibility of the sample is strongly dependent upon the sample shape. Due to the annular geometry of our powder enclosure, the 12 mm diameter solid cylindrical TiZr sample used to normalize the detectors can sometimes result in unusual nonphysical scattering intensities. To remedy this, we note that the total spin-flip cross section is $\sigma_{SF}=\sigma^x_{SF}+\sigma^y_{SF}+\sigma^z_{SF}=2\sigma_i + \sigma_m$, where $\sigma_i$ originates from spin-incoherent scattering and $\sigma_m$ is magnetic. In our case, we make the assumption that the elastic scattering is completely dominated by spin-incoherent scattering and there is no contribution from magnetism. This contrasts with the inelastic channel where we expect the dominant spin-flip scattering to be magnetic. Then, the elastic line should be completely Q-independent, and one can simply normalize the detectors to this quantity. In this way, one can neatly avoid the issues regarding the sample visibility by normalizing the scattering to the sample itself. A comparison of this resulting normalization to the standard TiZr normalization is shown in Fig. \ref{fig:HYS_norm}, which is very similar as one would expect. 

\begin{figure}
    \centering
    \includegraphics[width=1\columnwidth]{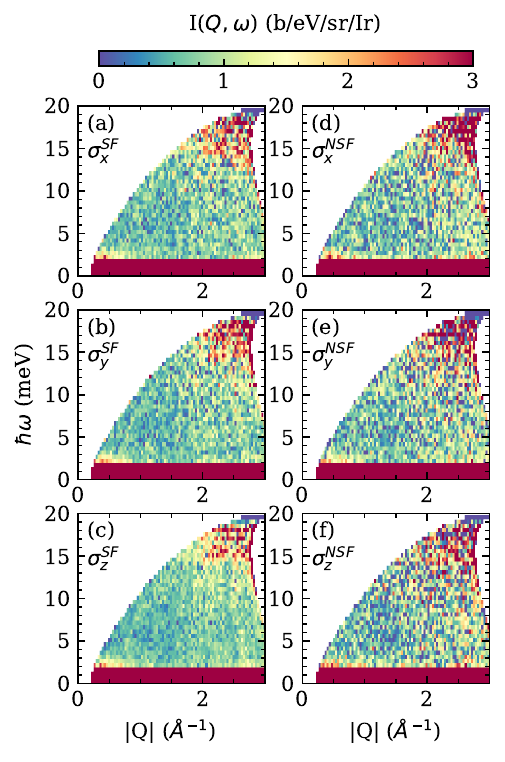}
    \caption{Individual scattering intensities from the six different polarization configurations in the HYSPEC measurement.}
    \label{fig:hys_individual_crosssections}
\end{figure}

Finally, all the individual cross sections may be compared, and relevant experimental quantities discussed in the main text may be extracted. The individual cross sections are all extremely similar, but are pictured for completeness in Fig. \ref{fig:hys_individual_crosssections}(a-f).

\section{Details of THz Spectroscopy measurements}

\begin{figure}[h]
    \centering
    \includegraphics[width=1.0\columnwidth]{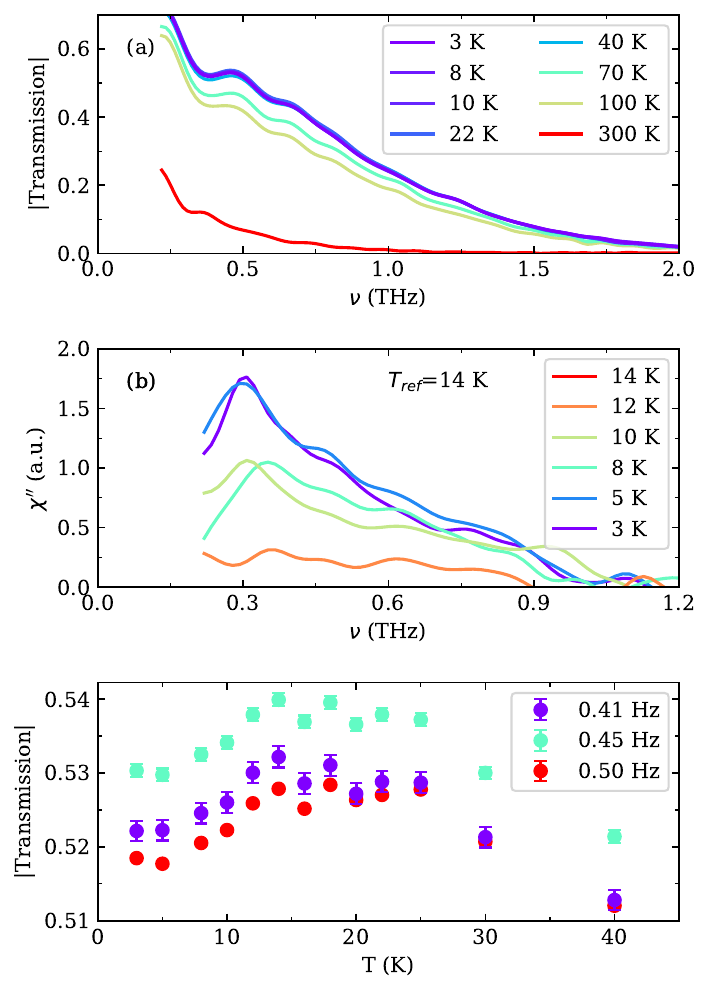}
    \caption{Summary of temperature dependent THz studies on pellets of D$_3$LiIr$_2$O$_6$. }
    \label{fig:THz_Tdep}
\end{figure}

Time-domain THz spectroscopy (TDTS) was conducted on two independent 5 mm diameter and 1.0(2) mm thick pressed pellets of $^3$D$^7$Li$^{193}$IrO$_3$ from the same growth as the neutron samples using time domain THz spectroscopy (TDTS). Measurements were performed using a custom-built system with frequency range 0.2-2~THz \cite{Laurita2017LowMagnets} at zero magnetic field. Using a non-magnetic reference signal, one may extract the dynamical susceptibility $\chi_M(\omega)$ at $Q=0\ \AA^{-1}$ which cannot be accessed in these INS measurements~\cite{Chauhan2020TunableChain, HalloranLi2IrO3}. This measurement is complicated by the presence of fluctuating H/D ions which introduce electric dipole moments within the material which strongly scatter the incoming THz light. Direct transmission spectra are shown in Fig. \ref{fig:THz_Tdep}(q).

By comparing the lowest available temperature of $T$=2.0(1) K and $T$=20.0(1) K measurements, a temperature-dependent clear continuum of excitations in the dynamical susceptibility is observed below $\hbar\omega=$5 meV extending down to the lowest accessible energy transfer near 1 meV consistent with the INS data. The measured complex transmission of the THz pulse is converted to a susceptibility $\chi''(\hbar\omega)$ by a referencing method in the same manner as in Ref.~\cite{HalloranLi2IrO3}. The reference temperature was chosen based upon the peak in temperature dependent THz transmission at the relevant energy scale around 0.4 THz, as shown in Fig.~\ref{fig:THz_Tdep}(c).

\bibliography{refs}